\documentclass[useAMS,usenatbib]{mn2e}
\usepackage{longtable}
\usepackage{epsfig}
\usepackage{amssymb}

\title[The dark matter haloes of Chandra X-ray galaxy clusters and baryons effect]{The dark matter haloes of Chandra X-ray galaxy clusters and baryons effect}

\author[Iu. Babyk, I. Vavilova, A. Del Popolo]{Iu. Babyk$^{1,4,5,6}$\thanks{E-mail:
babikyura@gmail.com}, I. Vavilova$^{4}$\thanks{E-mail: iriv@mao.kiev.ua}, A. Del Popolo$^{2,3}$\thanks{E-mail:
antonino.delpopolo@unibg.it}\\
$^{1}$Faculty of Physics of National Taras Shevchenko University of Kyiv, Glushkova ave., 2, 03127, Kyiv, Ukraine\\
$^{2}$Dipartimento di Fisica e Astronomia, Universit\'a di Catania, Viale Andrea Doria 6, 95125 Catania, Italy\\
$^{3}$ Departamento de Astronomia, Universidade de S\~ao Paulo, Rua do Mat\~ao 1226, 05508-900, S\~ao Paulo, SP, Brazil\\
$^{4}$ Main Astronomical Observatory NAS of Ukraine, Zabolotnogo str., 27, 03650, Kyiv, Ukraine\\
$^{5}$ Dublin Institute for Advanced Studies, 31 Fitzwilliam Place, Dublin 2, Ireland \\
$^{6}$ Dublin City University, Dublin 9, Ireland \\
}

\begin{document}

\date{Accepted \underline{\hspace*{2cm}}. Received \underline{\hspace*{2cm}}; in original form \underline{\hspace*{2cm}}}

\pagerange{\pageref{firstpage}--\pageref{lastpage}} \pubyear{0000}

\maketitle

\label{firstpage}

\begin{abstract}
We present results based on Chandra observations of a large sample of 129 hot galaxy clusters. We measure the concentration parameter c$_{200}$, the dark mass M$_{200}$ and the baryonic mass content in all the objects of our sample, providing the largest dataset of mass parameters for galaxy clusters in the redshift range $z$ = 0.01 -- 1.4. We confirm that a tight correlation between c$_{200}$ and M$_{200}$, $c \propto M^a_{vir}/(1+z)^b$ with $a = $ -0.56 $\pm$ 0.15 and $b = $0.80$ \pm$ 0.25 (68 per cent confidence limits), is present, in good agreement with the predictions from numerical simulations and previous observations. Fitting the mass profile with a generalized NFW model, we got the inner slope $\alpha$, with $\alpha = 0.94 \pm 0.13$.
Finally, we show that the inner slope of the density profile, $\alpha$ correlates with the baryonic mass content, $M_b$: namely  $\alpha$ is decreasing with increasing baryonic mass content.
\end{abstract}


\section{Introduction}\label{sec:1}

\indent\indent 

The problem of dark matter halos formation has a long history going back to the seminal paper of \citet{Gunn:72} and \citet{Gunn:77}, who studied the density profile formation using the collapse of a spherical perturbation in an expanding background. The quoted papers and several following analytical models (e.g., \citealt{Fillmore:84}; \citealt{Bertschinger:85}; \citealt{Hoffman:85}), found that density profiles are described by power-laws in all the radius range. More recent semi-analytical models (e.g., \citealt{White:92}, \citealt{Subramanian:00}; \citealt{Delpopolo:00}; \citealt{El-Zant:01}, \citealt{El-Zant:04}; \citealt{Hiotelis:02}; \citealt{LeDelliou:03}; \citealt{Ascasibar:04}; \citealt{Williams:04}; \citealt{Tonini:06}; \citealt{Ascasibar:07}) showed that the profile is not a power-law, similarly to N-body simulations results (e.g., \citet{Navarro:96}, \citet{Navarro:97}); 
Navarro--Frenk--White (NFW)), \citealt{Moore:98}; \citealt{Jing:00}; \citealt{Klypin:01}, \citealt{Bullock:01}, \citealt{Power:03}, and \citealt{Navarro:04}, \citealt{Navarro:10}) found that the spherically averaged density profiles of the N-body DM halos are similar, regardless of the mass of the halo or the cosmological model.
The NFW profile is given by:
\begin{equation}
\rho(r)= \frac{\rho_0}{r/r_s(1+r/r_s)^2}=\frac{\rho_{critic} \Delta_c }{r/r_s(1+r/r_s)^2}
\end{equation}
where $\rho_{critic}$ is the critical density of the Universe at the cluster's redshift $z$, and $\Delta_c $ is the virial overdensity.
The scale radius $r_s$ is connected to the virial radius $r_{vir}$ through the concentration parameter $c$, $c= r_{vir}/r_s$\footnote{Since $r_{vir}$ is difficult to determine observationally, its value is often approximated by the radius within which the average density is greater than the critical density by a specified factor (e.g., 200). In the following of the paper, $r_{vir}$ is identified with $r_{200}$.}. 
At small scales, the logarithmic density slope, that is also known as inner slope, is given by
\begin{equation}
\alpha= -\frac{d \log{\rho}}{d  \log{r}}|_{r \rightarrow 0}=1
\end{equation}
The quoted profile diverges as $\rho \propto r^{-1}$ in the inner part, and at large radii behaves as $\rho \propto r^{-3}$. The inner slope is even steeper in \citet{Moore:98} profile ($\rho \propto r^{-1.5}$). More recent simulations (e.g., \citealt{Power:03}, \citealt{Hayashi:04}, and \citealt{Navarro:04}, \citealt{Navarro:10}; \citealt{Stadel:09}) showed that density profiles are better fitted by the Einasto profile, which becomes shallower towards the centre of the halo. 

Unfortunately, N-body simulations predictions are not supported by observations. Observations of the inner part of density profiles of dwarfs galaxies and LSBs are characterized by a core-like structure (e.g. \citealt{Flores:94}; \citealt{Moore:94}; \citealt{deBlok:02}; \citealt{deBlok:03}; \citealt{Gentile:04}, \citealt{Salucci:07}; \citealt{deNaray:08}, \citealt{deNaray:09}), and a similar problem is evidenced when studying the clusters of galaxies inner profile.
Density profiles of clusters has been studied through X-ray observations, strong and weak lensing.  
X-ray temperature measurements give information on cluster structure in the range 500-50 kpc (\citealt{Brada:08}). At smaller radii, temperature determination is limited by instrumental resolution or substructure (\citealt{Schmidt:07}, hereafter SA07). X-ray measurements are also limited by ``cooling flows" presence, and the breaking of assumption of hydrostatic equilibrium (see \citealt{Arabadjis:04}). The inner slope calculation through the use of X-ray observations brought to discrepant values ( e.g., 0.6 (\citealt{Ettori:02}), 1.2 (\citealt{Lewis:03}), 1.9 (\citealt{Arabadjis:02}). 

Another technique used to study the DM distribution in clusters is gravitational lensing. Weak lensing of background galaxies is used to reconstruct the mass distribution in the outer parts of clusters (\citealt{Mellier:99}). The resolution that can be achieved is able to constrain profiles inside~100 kpc. Strong lensing is used to study the DM distribution in the inner parts of clusters, it has a typical sensitivity to the projected mass distribution inside $\simeq 100-200$ kpc, with limits at $\simeq 10-20$ kpc (\citealt{Gavazzi:05}; \citealt{Limousin:08}, Alexandrov et al. 2011; Tsvetkova et al. 2009). 

Discrepant results have been sometime obtained when using the lensing method. \citet{Smith:01}, found $\alpha >1$ at 1\% of $r_{vir}$ by studying the tangential and radial arcs of A383. A much smaller value $\alpha$ was obtained by \citet{Sand:04} and \citet{Newman:11}, for the same cluster, using lensing and through the aid of stellar kinematics of the central region. \citet{Tyson:98} found $\alpha=0.57 \pm 0.024$ for Cl 0024+1654, while  \citet{Kneib:03} found that a NFW profile fits the profile in the radius range 0.1-several $r_{vir}$. 
\citet{Sand:02} found a cored profile with $\alpha =0.35$ for MS2137.3-2353, while  \citet{Gavazzi:03} and \citet{Gavazzi:05} concluded that the precise value of the slope depends on the mass-to-light ratio of the Brightest Cluster Galaxy (BCG).  

In summary, N-body simulations are not always in agreement with the inner slopes of dwarf galaxies, LSBs and clusters of galaxies, and in the case of clusters of galaxies, observations may even disagree for the very same cluster (\citealt{Gavazzi:05} ; \citealt{Smith:05}; \citealt{Zappacosta:06}; SA07; \citealt{Brada:08}; \citealt{Limousin:08}; \citealt{Umetsu:08}). 

Several could be the reasons of the discrepancy (e.g., (a) different definition of the slope, which sometimes refers to
the DM and sometimes to the total mass; (b) use of observational techniques with different/limited dynamic
range in radius; (c) not taking into account the stellar mass of the BCG). In order to obtain more sure constraints on the central part of the density profiles, one should use combined methods (\citealt{Miralda-Escude:95}; \citealt{Kneib:03} (weak+strong lensing); \citealt{Brada:05} (weak+strong lensing); \citealt{Mahdavi:07} (X-ray+weak lensing)), or better constraints on the central part of the density profiles can be obtained through stellar kinematics of the central galaxy ($\simeq 1-200$ kpc region). 

In a series of papers, \citet{Sand:02}, \citet{Sand:04}, \citet{Sand:08}, studied the clusters MS 2137-23; A383; A963; RXJ1133; MACS 1206; A1201, separating the contribution to the halo coming from the DM from that coming from the baryonic and stellar mass of the BCG. They found a profile flatter than $\alpha <1$ except for RXJ1133. \citet{Newman:11} found $\alpha <1$ (95 per cent CL) for A383. \citet{Newman:09} presented a detailed analysis of DM and baryonic distributions in A611, finding a slope $\alpha <0.3$ (68 per cent CL). 

Summarizing, at least some clusters of galaxies have inner density-profile slopes shallower that those obtained in N-body simulations, in agreement with what happens, as previously reported, with the density profiles of dwarfs galaxies and LSBs (e.g. \citealt{Flores:94}; \citealt{Moore:94}; \citealt{deBlok:02}; \citealt{deBlok:03}; \citealt{Gentile:04}, \citealt{Salucci:07}; \citealt{deNaray:08}, \citealt{deNaray:09}). We want to recall here that, similarly to Sand's result concerning the quoted clusters, the galaxies NGC 2976, NGC 4605, NGC 5949, NGC5963 and NGC 6689, have inner slopes going from very flat to cuspy, and \citet{deBlok:08}, using a sample from The HI Nearby Galaxy Survey (THINGS) found that the best fit to rotation curves, and then the inner slope $\alpha$ of their density profile, depends on their mass. 

Several papers have shown the fundamental role of baryons in shaping the density profiles of structures. Different processes have been pointed out capable of flattening the inner density profile, transferring energy from stellar baryons to the dark matter, heating it and lowering the central dark matter density (\citealt{Milosavljevic:01}; \citealt{El-Zant:01}, \citealt{El-Zant:01}; \citealt{Weinberg:02}; \citealt{Loeb:03}; \citealt{Gao:07}; \citealt{Ma:09}; \citealt{McMillan:05};   \citealt{Tonini:06}; \citealt{Mashchenko:06}; \citealt{Romano-Diaz:08}, \citealt{Romano-Diaz:09}; \citealt{delPopolo:09}; \citealt{Governato:10}; Kulinich et al. 2012).

In the present paper, we use Chandra X-ray data to study, similarly to SA07, the properties of 129 dynamically relaxed galaxy clusters. We selected the sample with redshift range from 0.01 to 1.4. with the aim to recover their total and gas mass profiles and analysis of the measured distribution of $c_{200}$, $M_{200}$ and baryonic mass content. 
We determine the baryons content of each cluster subtracting the DM from the total mass. Even if the results concerning the density profile of clusters are similar to those of SA07, we point out that fitting the density profile with a generalized NFW model 
\begin{equation}
\rho(r)= \frac{\rho_0}{(r/r_s)^{\alpha}(1+r/r_s)^{3-\alpha}}
\end{equation}
the slope $\alpha$ is correlated with the baryonic mass content of the cluster, in agreement with several studies (\citealt{Ricotti:03}; \citealt{Subramanian:00}; \citealt{Simon:05}; \citealt{Cen:04}; \citealt{Ricotti:04}; \citealt{Williams:04}; \citealt{Ricotti:07}; \citealt{delPopolo:09}, \citealt{delPopolo:12}).
In the paper the assumed value of Hubble constant is $H_{0}=73 km s^{-1} Mpc^{-1}$ and $\Omega_{m}=0.27$.
The outline of our work is the following. In Section 2, we describe the X-ray galaxy clusters sample of Chandra observations compiled by us  to recover the baryonic and total mass profiles with techniques presented in Section 3. In Section 4, we present a preliminary discussion of the main results. We investigate the $c_{200}-M_{200}$ and $\alpha-M_{b}$ relations. We summarize our results and draw the conclusion of the present study in Section 5 and 6.

\section{Sample and Data processing}

\indent\indent  The sample has been built to cover a wide range in redshift from 0.01 to 1.4, constituted by 129 galaxy clusters. All clusters in the sample have a regular X-ray morphology, indicative of a relaxed state and allowing reliable determination of the total mass profile through the hydrostatic equilibrium equation. The observations results are presented in Tab. \ref{tab1}.
The preliminary processing of the Chandra data, which where obtained using the Advanced CCD Imaging
Spectrometer (ACIS), included: the Chandra pipeline processing of the list of events; the re-processing with the
CIAO software package; improving the rejection of cosmic ray events by VFAINT mode; cleaning the data in format of
recommendation by the Chandra X-ray Centre (to remove periods of anomalously high background).

\begin{table*}
\centering
\fontsize{8}{8}
\selectfont
\caption{\fontsize{9}{9} \selectfont Summary of the Chandra observations. Column list the target name, redshift, number of ID observation, net exposure time, detector used, column density and coordinates (taken from NED) from the X-ray centres used in the analysis.}
\bigskip
\label{tab1}
\begin{tabular}{llcccccc}
\hline
Cluster  & $z$  & $ObsID$  & $t_{exp}$, & Instrument & $n_{H}$, & RA(J2000) & DEC(J2000) \\
&&&ks&&10$^{20}$ cm$^{-2}$&&\\
\hline
A85&0.055&904&38.91&ACIS-I&3.26&00 41 37.8 &-09 20 33\\
A119&0.044&7918&45.63&ACIS-I&3.44&00 56 21.4 &-01 15 47\\
A133&0.056&2203&35.91&ACIS-S&1.55& 01 02 39.0 &-21 57 15\\
A168&0.045&3203&41.12&ACIS-I&3.48 &01 15 09.8 &+00 14 51\\
A209&0.206&522&10.09&ACIS-I&1.71&01 31 53.0 &-13 36 42\\
A262&0.016&2215&29.12&ACIS-S&5.36&01 52 50.4 & +36 08 46\\
A383&0.187&524&10.09&ACIS-I&3.84& 02 48 06.9 &-03 29 32\\
A399&0.072&3230&49.28&ACIS-I&11.50&02 57 56.4 & +13 00 59\\
A401&0.074&518&18.24&ACIS-I&10.30&02 58 56.9 &+13 34 56\\
A478&0.088&1669&42.94&ACIS-S&14.80&04 13 20.7 & +10 28 35\\
A496&0.033&931&19.16&ACIS-S&4.45&04 33 37.1 &-13 14 46\\
A521&0.250&430&39.60& ACIS-S & 6.05&04 54 10.1 &-10 15 11\\
A539&0.028&5808&24.60&ACIS-I&12.90&05 16 35.1 &+06 27 14\\
A576&0.039&3289&39.10&ACIS-S&5.66& 07 21 24.1 &+55 44 20\\
A644&0.070&2211&30.08&ACIS-I&6.44& 08 17 26.4 &-07 35 21\\
A697&0.280&4217&19.70&ACIS-I&3.41& 08 42 53.3 &+36 20 12\\
A754&0.054&577&44.77&ACIS-I&4.45& 09 08 50.1 &-09 38 12\\
A773&0.217&533&10.50&ACIS-I&	1.44& 09 17 59.4 &+51 42 23\\
A780&0.054&575&24.10&ACIS-I&4.87& 09 18 30.3 &-12 15 40\\
A907 &0.150&3205&47.70&ACIS-I&5.46& 09 58 21.1 &-11 03 22\\
A963&0.206&903&36.76&ACIS-S&1.40& 10 17 13.9 &+39 01 31\\
A1060&0.012&2220&32.32&ACIS-I&4.79&10 36 51.3 &-27 31 35\\
A1068&0.138&1652&27.17&ACIS-S&1.39& 10 40 47.1 &+39 57 19\\
A1201&0.169&4216&40.17&ACIS-S&1.66&11 13 01.1 &+13 25 40\\
A1240&0.159&4961&52.03&ACIS-I&1.98&11 23 32.1  &+43 06 32\\
A1361&0.117&2200&16.96&ACIS-S&2.23& 11 43 45.1& +46 21 21\\
A1413 &0.140&537&9.34&ACIS-I&2.05&  11 55 18.9 &+23 24 31 \\
A1446&0.104&4975&59.12&ACIS-S&1.41& 12 01 51.5 &+58 01 18\\
A1569&0.074&6100&41.75&ACIS-I&2.22&12 36 18.7 & +16 35 30\\
A1644&0.047&2206&18.96&ACIS-I&4.97& 12 57 14.8  &-17 21 13\\
A1650&0.084&5823&40.13&ACIS-I&1.57& 12 58 46.2 &-01 45 11\\
A1651&0.085&4185&9.77& ACIS-I&1.88&12 59 22.9  &-04 11 10\\
A1664&0.128&1648&9.90&ACIS-S&8.91&13 03 41.8 &-24 13 06\\
A1689 &0.183&540&10.45&ACIS-I&1.81&13 11 29.5 &-01 20 17\\
A1736&0.046&4186&15.12&ACIS-I&5.38& 13 26 52.1 &-27 06 33\\
A1795&0.063&493&19.88&ACIS-S&1.20&13 49 00.5 &+26 35 07\\
A1835&0.253&495&19.77&ACIS-S&2.30&14 01 02.0 &+02 51 32\\
A1914&0.171&542&8.17&ACIS-I&0.93&14 26 03.0 &+37 49 32\\
A1995&0.318&906&58.23&ACIS-S&1.45&  14 52 50.4 &+58 02 48\\
A2029&0.077&891&20.07&ACIS-S&3.07&15 10 56.0 &+05 44 41\\
A2034&0.113&2204&54.70&ACIS-I&1.59&15 10 13.1 &+33 31 41\\
A2052&0.035&890&37.23&ACIS-S&2.78&15 16 45.5 &+07 00 01\\
A2063&0.035&6263&17.04&ACIS-S&2.98&15 23 01.8 &+08 38 22\\
A2065&0.073&3182&50.09&ACIS-I&2.94&15 22 42.6 &+27 43 21\\
A2124&0.065&3238&19.61&ACIS-S&1.68&15 44 59.3 &+36 03 40\\
A2142&0.091&1228&12.26&ACIS-S&4.25&15 58 16.1 &+27 13 29\\
A2147&0.035&3211&18.12&ACIS-I&3.40& 16 02 17.2 &+15 53 43\\
A2163&0.203&545&9.57&ACIS-I&12.10&16 15 34.1 &-06 07 26\\
A2199&0.030&497&19.72&ACIS-S&0.87&16 28 38.5 &+39 33 06\\
A2204&0.152&499&10.20&ACIS-S&5.66&16 32 45.7 &+05 34 43\\
A2218&0.175&553&5.96&ACIS-I&3.30&16 35 54.0  &+66 13 00\\
A2219&0.225&896&42.84&ACIS-S&1.75&16 40 21.1 &+46 41 16\\
A2244&0.097&4179&57.72&ACIS-S&2.10&17 02 44.0 &+34 02 48\\
A2256&0.058&519&15.00&ACIS-I&4.11&17 03 43.5 &+78 43 03\\
A2319&0.056&3231&14.62&ACIS-I&7.85&19 20 45.3 &+43 57 43\\
A2390&0.228&500&9.96&ACIS-S&6.94&21 53 34.6 &+17 40 11\\
A2462&0.073&4159&39.74&ACIS-S&3.11&22 39 05.2 &-17 21 22\\
A2537&0.295&4962&36.67&ACIS-S&4.50&23 08 16.4 &-02 10 44\\
A2589&0.041&3210&13.86&ACIS-S&4.15&23 24 00.5 &+16 49 29\\
A2597&0.085&922&39.86&ACIS-S&2.50&23 25 18.0 &-12 06 30\\
A2634&0.031&4816&50.16&ACIS-S&5.02&23 38 18.4 &+27 01 37\\
A2657&0.040&4941&16.36&ACIS-I&5.86&23 44 51.0 &+09 08 40\\
A2667&0.230&2214&9.77&ACIS-S&1.64&23 51 47.1 &-26 00 18\\
A2670&0.076&4959&40.14&ACIS-I&2.91&23 54 13.7 &-10 25 08\\
A2717&0.049&6973&47.65&ACIS-I&1.12&00 02 59.4 &-36 02 06\\
A2744 &0.308&2212&25.14&ACIS-S&1.60&00 14 19.5 &-30 23 19\\
A3112&0.075&6972&30.16&ACIS-I&2.72&03 17 52.4 &-44 14 35\\
A3158&0.060&3712&31.35&ACIS-I&1.62&03 42 39.6 &-53 37 50\\
A3266&0.059&899&30.14&ACIS-I&1.71&04 31 11.9 &-61 24 23\\
A3376&0.045&3202&44.85&ACIS-I&4.71&06 00 43.6 &-40 03 00\\
A3391&0.051&4943&18.69&ACIS-I&5.58&06 26 15.4 &-53 40 52\\
A3395&0.050&4944&22.17&ACIS-I&6.13&06 27 31.1 &-54 23 58\\
A3526&0.011&504&32.12&ACIS-S&8.07&12 48 51.8 &-41 18 21\\
A3558&0.048&1646&14.61&ACIS-S&3.89&13 27 54.8 &-31 29 32\\
A3562&0.049&4167&19.54&ACIS-I&3.83&13 33 31.8 &-31 40 23\\
\hline
\end{tabular}
\end{table*}

\setcounter{table}{0}
\begin{table*}
\centering
\fontsize{8}{8}
\selectfont
\caption{\fontsize{9}{9} \selectfont Continued.}
\bigskip
\begin{tabular}{llcccccc}
\hline
Cluster  & $z$  & $ObsID$  & $t_{exp}$, & Instrument & $n_{H}$,&RA(J2000)&DEC(2000)\\
&&&ks&&10$^{20}$ cm$^{-2}$&&\\
\hline
A3571&0.039&4203&34.44&ACIS-S&3.91&13 47 28.9 &-32 51 57\\
A3667&0.055&889&50.96&ACIS-I&4.95&20 12 30.1 &-56 49 00\\
A3827&0.098&7920&46.17&ACIS-S&2.12&22 01 49.1 &-59 57 15\\
A3921&0.093&4973&29.76&ACIS-I&2.82&22 49 38.6 &-64 23 15\\
A4038&0.030&4992&33.97&ACIS-I&1.54&23 47 37.0 &-28 07 42\\
A4059&0.047&897&41.19&ACIS-S&1.10&23 56 40.7  &-34 40 18\\
AWM4&0.032&9423&75.46&ACIS-S&5.03&16 04 57.0 &+23 55 14\\
CLJ1226.9+3332&0.890&3180&32.12&ACIS-I&1.37&12 26 58.0 &+33 32 54\\
Coma (A1656)&0.023&10672&28.91&ACIS-S&0.93&12 59 48.7 &+27 58 50\\
IIIZw54&0.0311&4182&23.76&ACIS-I&15.50&03 41 17.6 &+15 23 44\\
ISCS J1438+3414 & 1.41 & 10461 & 150.00 & ACIS-S& 2.01& 14 33 22.4 & +22 18 35.4 \\
MACSJ0011.7-1523 &0.379&3261&22.00&ACIS-I&2.05&00 11 42.9 &-15 23 22\\
MACSJ0159.8-0849 & 0.405&3265&18.00&ACIS-I&2.08&01 59 48.0 &-08 49 00\\
MACSJ0242.6-2132 &0.314&3266&12.00&ACIS-I&2.69&02 42 36.0 &-21 32 00\\
MACSJ0429.6-0253 &0.399&3271&24.00&ACIS-I&5.25&04 29 36.0 &-02 53 00\\
MACSJ0647.7+7015&0.59&3196&18.85&ACIS-I&5.63&06 47 45.9 &+70 15 03\\
MACSJ0744.8+3927&0.697&536&19.69&ACIS-I&5.68&07 44 51.8 &+39 27 33\\
MACSJ1115.8+0129 &0.355&3275&16.00&ACIS-I&4.39&11 15 53.3 &+01 29 47\\
MACSJ1311.0-0311 &0.49&3258&15.00&ACIS-I&1.88&13 11 00.0 &-03 11 00\\
MACSJ1423.8+2404 &0.545&4195&113.40&ACIS-I&2.38&14 23 48.3 &+24 04 47\\
MACSJ1427.6-2521 &0.318&3279&17.00&ACIS-I&6.18&14 27 39.4 &-25 21 02\\
MACSJ1720.3+3536 &0.391&3280&21.00&ACIS-I&3.37& 17 20 15.5 &+35 36 21\\
MACSJ1931.8-2635 &0.352&3282&14.00&ACIS-I&9.11&19 31 48.0 &-26 35 00\\
MACSJ2129.4-0741&0.589&3199&17.69&ACIS-I&4.84& 21 29 26.0 & -07 41 28\\
MACSJ2229.8-2756 &0.324&3286&17.00&ACIS-I&1.34&22 29 48.0 &-27 56 00\\
MKW3s&0.045&900&58.03&ACIS-I&3.04&15 21 51.9 &+07 42 31\\
MKW4&0.020&3234&30.36&ACIS-S&1.90&12 03 57.7 &+01 53 18\\
MKW8 &0.027&4942&23.45&ACIS-I&2.80&14 40 43.1 &+03 27 11\\
PKS0745-19&0.103&508&28.33&ACIS-S&42.70&07 47 31.32& -19 17 40.0\\
RXCJ0043.4-2037 &0.292&9409&20.18&ACIS-I&1.54&00 43 23.1 &-20 37 35\\
RXCJ0232.2-4420 &0.283&4993&23.71&ACIS-I&2.42&02 32 18.7 &-44 20 41\\
RXCJ0307.0-2840 &0.253&9414&19.16&ACIS-I&1.36&03 07 01.1 &-28 40 30\\
RXCJ0516.7-5430 &0.294&9331&9.64&ACIS-I&6.80&05 16 35.2 &-54 16 37\\
RXCJ0547.6-3152 &0.148&9419&20.04&ACIS-I&2.02&05 47 34.2 &-31 53 01\\
RXCJ0605.8-3518 &0.141&12899&5.06&ACIS-I&4.47&06 05 52.4 &-35 18 22\\
RXCJ1131.9-1955 &0.307&3276&14.10&ACIS-I&4.46&11 32 00.7 &-19 53 34\\
RXCJ2014.8-2430 &0.161&11757&20.18&ACIS-S&7.77&20 14 49.7 &-24 30 30\\
RXCJ2129.6+0005 &0.235&552&10.09&ACIS-I&4.29&21 29 37.9 &+00 05 39\\
RXCJ2337.6+0016 &0.273&3248&9.31&ACIS-I&3.82&23 37 39.7 &+00 17 37\\
ZwCL1215&0.075&4184&12.22&ACIS-I&1.74&12 17 40.60& +03 39 45.0\\
RCSJ0224-0002&0.778&4987&90.15&ACIS-S&2.92&02 24 00.0 &-00 02 00\\
RCSJ0439-2904&0.951&3577&77.17&ACIS-S&2.63&04 39 38.0 &-29 04 55\\
RCSJ1107-0523&0.735&5825&50.12&ACIS-S&4.24&11 07 22.80& -05 23 49.0\\
RCSJ1419.2+5326&0.64&3240&10.03&ACIS-S&1.18&14 19 12.0 &+53 26 00\\
RCSJ1620+2929&0.87&3241&37.13&ACIS-S&2.72&16 20 09.40& +29 29 26.0\\
RCSJ2156+0123&0.335&674&45.76&ACIS-S&2.56&15 47 34.2 & +26 38 29.0\\
RCSJ2318+0034&0.78&4938&51.12&ACIS-S&4.13&23 18 30.67 & +00 34 03.0\\
RCSJ2319+0038&0.904&5750&21.18&ACIS-S&4.16 &23 19 53.00& +00 38 00.0\\
RXJ0439.0+0520&0.208&527&9.71&ACIS-I&10.50&04 39 02.2 &+05 20 43\\
RXJ0848.7+4456&0.570&927&126.74&ACIS-I&2.66& 08 48 47.2 &+44 56 17\\
RXJ0849+4452&1.26&945&128.45&ACIS-I&2.50&08 53 43.6 & +35 45 53.8\\
RXJ0910+5422&1.106&2227&84.20&ACIS-I&2.35&09 10 45.36&+54 22 07.3\\
RXJ1113.1-2615&0.730&915&105.95&ACIS-I&5.47&11 13 05.2 &-26 15 26 \\
RXJ1221.4+4918 & 0.700& 1662& 80.13& ACIS-I& 1.45& 12 21 24.5 &+49 18 13\\
\hline
\end{tabular}
\end{table*}

The Tab. 1 consists of the name of clusters, the redshift, the number of observation, the exposition time (before cleaning), instrument, and the values of column density, which were described by \citet{Dickey:90}, and coordinates (RA, DEC) (which were taken from NED\footnote{http://nedwww.ipac.caltech.edu}). It is important to note that our list of clusters includes a few clusters with merger's properties, for example, Abell 2744, but in such case we have used only one component on the image for data reduction.
\begin{figure}
\centering
\epsfig{file=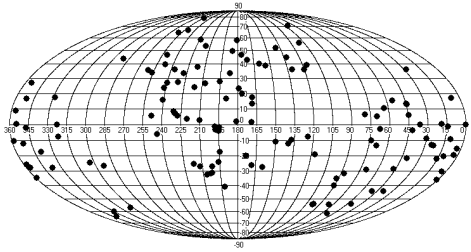, width=0.8\linewidth}
\caption{The distribution of galaxy clusters at the celestial sphere.}\label{graph}
\end{figure}

\begin{figure}
\centering
\epsfig{file=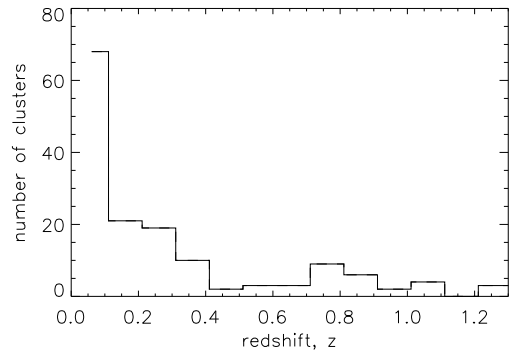, width=0.8\linewidth}
\caption{The distribution of galaxy clusters in our sample.}\label{figure}
\end{figure}

As we see in Fig.\ref{figure}, the most of clusters has the redshift in the range 0.01 -- 0.3. We added to our sample MACS and RCSJ clusters as ones of the most distant objects which are observed by Chandra. The distribution of the studied galaxy clusters at the celestial sphere is given in Fig. \ref{graph}. We used in our research the sample of the clusters used by \citet{Allen:02} to study the evolution of the X-ray gas mass fraction and constrain cosmological parameters. To minimize systematic scatter and to allow the most precise test of the CDM model predictions, we used only highly relaxed clusters.

The data reduction was performed with the Chandra CIAO v.4.2 including CalDB 4.5.1 for maps and calibration.
The main data processing steps were provided using techniques discussed by \citet{Babyk:12a}, \citet{Babyk:12b}, 
\citet{Babyk:12c}, \citet{Babyk:12d} in our previous research. The quoted steps are the following: 
a) removing the point sources; b) determining the X-ray peaks; c) centering the X-ray peacks with different width to provide the  same number of X-ray photons in each annulus; d) extracting the annular spectra.
The average of the outermost radius is about 1 Mpc. We have generated ARF and RMF files using the mkarf and
mkrmf command in CIAO tool.

We used Xspec (version 12.6)(Arnaud 1996) to analyze spectra, the MEKAL\footnote{MEKAL is the model which describes an emission from hot diffuse plasma (ICM)} code (Kaastra \& Mewe 1993) and WABS\footnote{The WABS is a parameter which describe the galactic absorption (\citealt{Dickey:90})} for fitting the data and Fe-L calculations (Liedahl et al. 1995). The content of heavy elements in this model was taken close to the solar level (Z=0.3) and was frozen. The energy range 0.5-7.0 keV was used during fitting, and the spectra were grouped to have at least 50 counts per spectral
bin.

Background spectra were extracted from the blank-field data sets which is available from the Chandra X-ray centre, and this technique was used only for nearer galaxy clusters (0$< z <$0.3). For more distant clusters, the background spectra were extracted from the region which has the same size like in region for source. Usually, we used observations which were made on ACIS chips 0,1,2,3 and 7, because these chips are most accurate in calibration, although ACIS 4,5 and 6 were also used. 
All background spectra were cleaned as done with the main previous spectra.

We have used DSDEPROJ routine method (\citealt{Sanders:07}, \citealt{Russell:08}) to process the annular
spectra in order to determine X-ray temperature and other parameteres. After determining the temperature of clusters
we built the surface brightness profile for each cluster. DSDEPROJ is the deprojection routine which assumes only
spherical geometry, and solves some of the issues inherent to model-dependent deprojection routines. DSDEPROJ produces
a set of ``deprojected spectra'' which can then be fitted by a suitable spectral model in Xspec. We have shown that
this method does not generate the oscillating temperature profiles for multi-temperature clusters and produces a stable solution for an elongated cluster and clusters with breaks in temperature or density.\footnote{The DSDEPROJ source code is available at \textbf{www-xray.ast.cam.ac.uk/papers/dsdeproj}.}

For each ring we determined the temperature, $kT$, and the parameter $norm \sim \int n_{e}n_{H}dV$ which is proportional to the electron, $n_e$ and hydrogen, $n_H$, concentrations. Note, that the other parameters of model were fixed. Using these parameters, we have built the surface brightness profiles which fitted our theoretical surface profiles and then it was used to obtain the mass of each cluster (see below). The surface brightness profiles are measured in the 0.5-7 keV energy band, which provides an optimal ratio of the cluster and background flux in Chandra data. 

Note, that for lower redshift clusters in our sample, the statistical accuracy of the surface brightness at large radii is limited mostly by the Chandra field of view.

\section{Methods}

\subsection{Galaxy clusters mass profile}

In order to calculate the total mass profiles of galaxy clusters, spherical symmetry and hydrostatic equilibrium of clusters was
assumed. Similarly to SA07, we modeled the total mass profile as composed of three components: the dark matter halo, the gaseous component and the optically luminous mass of the central cD galaxy.

For the density profile of DM, we used the NFW model having a DM matter content
within a fixed radius given by
 \begin{equation}
 \label{mass_nfw} M(<r) = 4\pi\rho_{0} r^{3}_{s}\left[ln(1+\frac{r}{r_{s}})-\frac{r}{r_{s}+r}\right].
 \end{equation}
The gravitational potential $\phi$ of dark matter can be calculated by the formula
\begin{equation}
  \label{poten_nfw} \frac{d\phi}{dr} = G\frac{M(<r)}{r^{2}}.
 \end{equation}
We made all the computations assuming the hydrostatic equilibrium condition of the X-ray emitting gas and wrote the hydrostatic equilibrium condition as follows
 \begin{equation}
\label{hydros_nfw} \bigtriangledown P = -\rho_{g}\bigtriangledown \phi(r),
\end{equation}
where $P$ and $\rho_{g}$ are the gas pressure and the density, respectively. Since the density and pressure of hot gas have very low values, we can use the ideal gas law, $P = \frac{\rho_{g}kT_{g}(r)}{\mu m_{p}}$, where $\mu$ is molecular weight, and $m_p$ the proton mass. We then obtained the equation for unknown gas density distribution
\begin{equation}
\label{sys_nfw} \frac{\bigtriangledown \rho_{g}}{\rho_{g}} = -\bigtriangledown \phi(r) \frac{\mu m_{p}}{kT_{g}(r)}.
\end{equation}
For the construction of a hot gas density scalar field of galaxy clusters, we had to calculate numerically the system of the following differential equations 
\begin{equation}
\label{diffura_nfw} \frac{1}{\rho_{g}} \frac{\partial \rho_{g}}{\partial x_{i}} = -\frac{\mu m_{p}}{kT_{g}(r)} \frac{\partial \phi(r)}{\partial x_{i}},
\end{equation}
 for gas distribution, where $x_{i} = (x,y,z)$ are the Cartesian coordinates, and then for total mass  
\begin{equation}
\label{12}
M(<r) = -(kT(r)r/G\mu m_{p})\left(\frac{dln \rho}{dln r}  + \frac{dln T}{dlnr}  \right).
\end{equation}

In case of a single spherically-symmetric galaxy clusters, it is possible to calculate gas density and surface brightness profile analytically, like in the works of \citet{Humphrey:06} and \citet{O'Sullivan:07}, for example. We performed the numerical integrations of 
Eq. (\ref{diffura_nfw}) and Eq. (\ref{12}) by means of the Euler method. 
The integration step was chosen in order to obtain the maximum error of resulting X-ray brightness not bigger than 0.1 per cent. 

\subsubsection{Modeling of hot gas emission}

After 3D reconstruction of gas density field of pair, we were able to obtain the surface brightness in X-ray energy band.  The emission flux, that comes from the observed regions, is proportional to the square of the integral density of hot gas, $\rho^{2}_{g}$, along the line of sight. We used the metal abundance of 0.3 
and took $n_{e}/n_{p} = 1.17$ and $\rho_{g} = 1.35m_{p}n_{p}$, where $n_{e}$ and $n_{p}$ are the electron and proton concentrations, respectively (\citet{Vikhlinin:99}). We have to note that from the model, we can obtain the gas density with precision of some constant factor $a$ which depends on the integration boundary conditions of  Eq.~(\ref{diffura_nfw}). Therefore the real density $\rho_{g}^{real}$ of gas is equal to $\rho_{g}^{real} = a\cdot \rho^{sim}_{g}$, where $\rho^{sim}_{g}$ is the density of gas obtained solving the previous differential equations, that we indicate with $sim$ standing for simulations. 
So, we can write $\rho^{sim}_{g}\equiv \rho_{g}$ in Eq. (\ref{diffura_nfw}). Thus the value of emission measure $EM_{sim}$ from simulations can be found as
\begin{equation}
 \label{EM_sim} EM_{sim} = \int n_{e} n_{p} dV = 0.64/m^{2}_{p} \int (\rho^{sim}_{g})^{2} dV.
\end{equation}
Using equation (\ref{EM_sim}), we found the next expression
\begin{equation}
 \label{norm_sim} norm_{sim} = EM_{sim}\cdot 10^{-14}/[4\pi(D_{A}(1+z))^{2}],
\end{equation}
where $D_{A}$ is the angular distance to cluster. From the observational data we got the normalization parameter of the best-fit MEKAL model $norm_{MEKAL}$ which was fitted by the $norm_{sim}$ values from the simulations. Since $norm_{MEKAL}$ parameter can be  expressed by
\begin{equation}
 \label{norm_mekal} norm_{MEKAL} = \frac{0.64}{m^{2}_{p}}  \int \frac{(\rho^{real}_{g})^{2} dV \cdot 10^{-14}}{[4\pi(D_{A}(1+z))^{2}]},
\end{equation}
from Eq. (\ref{EM_sim}, \ref{norm_sim}) and Eq. (\ref{norm_mekal}), we have found the normalization factor $a$ as
\begin{equation}
 \label{a} a = \sqrt{\frac{norm_{MEKAL}}{norm_{sim}}}.
\end{equation}
Therefore we able to obtain the field of hot gas density $\rho^{real}_{g} = a\cdot \rho^{sim}_{g}$ and emission measure $EM_{real}$ putting into \ref{EM_sim} $\rho^{real}_{g}$ instead of $\rho^{sim}_{g}$. 

For each cluster we made the series of numerical simulations of the surface brightness profile reconstruction. In our model we used two free parameters $\rho_{0}$ and $r_{s}$ from NFW model. Sorting them out in the acceptable ranges, we selected such pair of values which is the best convenient for describing the observational cluster profile. Using the $\chi^{2}$ test we found the area of values $\rho_{0}$ and $r_{s}$ with 68\% confidence level. 

\subsubsection{Stellar component}


Our sample includes clusters, which have a single, optically dominant cD galaxy near their centers, and the mass of the stars in the central galaxy was accounted through the Jaffe model (\citealt{Jaffe:83}) with same parameters as in SA07.  
We assumed, as SA07 a mass of 1.14 $\times$ 10$^{12}$ M$_{\odot}$ for the stellar mass of each central galaxy in the sample.

\subsubsection{Best-fit 
}

All computations were made using $\chi^{2}$-test
\begin{equation}
\chi^{2} = \sum_{N} (\frac{V_{obs}-V_{model}}{\sigma_{obs}})^{2},
\end{equation}
where $V_{obs}$ and $V_{model}$ are the observed and modeled values, respectively. 
The regions with residual substructure, noticed by SA07, were down-weighted in the mass analysis, similarly to SA07. 

\subsubsection{Mass profiles scaling}

Density profiles of DM are approximatively universal on a large mass scale (Navarro, Frenk, \& White 1995; Navarro, Frenk, \& White 1996).\footnote{As previously reported, more recent simulations (e.g., \citealt{Power:03}, \citealt{Hayashi:04}, and \citealt{Navarro:04}, \citealt{Navarro:04}; \citealt{Stadel:09}) showed that density profiles are better fitted by the Einasto profile.}

In our work we scaled mass profiles with $R_{200}$ and $M_{200}$ (here $R_{200}$ is the radius within which the mean halo density
is 200 times the critical density, and $M_{200}$ is the total mass of galaxy cluster inside a sphere with radius $R_{200}$). 

The total mass within a given overdensity $\Delta$ is defined in the present work as 
\begin{equation}
\label{mass1} M_{\Delta} = \frac{4}{3}\pi \Delta \rho_{critic} R_{\Delta}^{3},
\end{equation}
where, as seen in the Introduction, R$_{\Delta}^{3}$ = $c_{\Delta} r_{s}$ is the radius within which the mean cluster overdensity is $\Delta$ times $\rho_{critic}$ and the relation with the concentration $c_{\Delta}$ and the scale radius $r_{s}$ holds by definition of the NFW mass profile. In the present paper, we assumed $\Delta=200$. 

An important issue that we are dealing with in this paper, is to understand if the baryonic content in the clusters has influences on the slope of the density profile. To this aim, for each object we calculated the fraction $M_b/M_{200}$. 
The quoted ratio $M_b/M_{200}$ will be obtained subtracting from the total mass the DM mass in each cluster.
So, the baryonic fraction will be derived as
\begin{equation}
\label{mass_b} M_{b} = \frac{800 \pi}{3} \rho_{critic} \left[R_{200,DM+b}^{3}-R_{200,DM}^{3} \right].
\end{equation}

We will then plot the obtained values of $\alpha$, namely the fit to DM component through a gNFW model, versus the baryonic content $M_b$.

\section{Results}

In this section we show and discuss the results for the total mass profiles, the DM profiles, the mass-concentration relation, and the inner slope of the DM profile. 

\subsection{Total mass profile}

In Tab. \ref{res1}, we summarized the results obtained from modeling the total mass profiles with a NFW  model ($\alpha=1$), similarly to SA07. We plotted just the result concerning the NFW fit and discard the singular isothermal sphere (SI) ($\rho(r) \propto r^{-2}$) since the result of the comparison to the total mass profile with a NFW mode and that with a SI is similar to SA07. The first column includes the names of galaxy cluster studied, the second column the parameter $r_s$, the third $c_{200} = R_{200}/r_s$, the fourth shows the value of $R_{200}$, the fifth includes  the value of total mass at $R_{200}$, and the sixth shows the goodness of fit ($\chi^2$/d.o.f.). Confidence limits are 68 per cent. For most clusters the NFW model is a good fit, similarly to SA07.

\subsection{Dark matter profile}
  
In Tab. \ref{res4}, we summarize the result from the analysis in which cluster mass distributions were separated in DM halo (fitted with a gNFW model), the X-ray emitting halo, and the luminous mass of the BCG. 

We considered as SA07 two cases. In the first case, we examined models in which the inner slope of the DM profile was fixed at $\alpha=1$,
corresponding to the NFW model, and in the second, we decomposed the total mass in its DM, diffused gas, and BCG, component (see Sect. 4.3). Again, except some cases, the NFW model provides a good fit to the DM profiles in clusters. 

We also calculated the concentration parameter for each cluster and this was used to obtain the mass-concentration parameter.
This was done calculating $c_{vir} \equiv c_{200}$ and $M_{vir} \equiv M_{200}$\footnote{Note that in SA07 $c_{vir}$, and 
$M_{vir}$ are connected through 
\begin{equation}
M_{vir}= \frac{4}{3} \pi r_{vir}^3 \Delta_c(z) \rho_{critic}
\end{equation}
\citet{Shaw:06}, with $\Delta_c = 178 \Omega_m(z)^{0.45}$ (\citealt{Lahav:91}) 
}
where 
\begin{equation}
\label{mass} M_{200} = \frac{4}{3}\pi 200 \rho_{critic} R_{200}^{3}
\end{equation}

\begin{table*}
\centering
\fontsize{7}{7}
\selectfont
\caption{\fontsize{9}{9} \selectfont The list of results from modeling the total mass profiles with Navarro-Frenk-White model with slope $\alpha$ = 1. First column includes names of galaxy cluster, second shows the $r_{s}$ parameter from NFW model, third shows the concentration parameter $c_{200}=R_{200}/R_{s}$ for the NFW model, fourth shows the radius within which the matter density is 200 times the critical density for total and dark matter profiles (in Mpc), fifth and sixth demonstrate the total mass in clusters at radius $R_{200}$ and $\chi^{2}/$d.o.f. goodness of fit. Confidence limits are 68 per cent.}
\bigskip
\label{res1}
\begin{tabular}{lccccc}
\hline
&&&&\\
\multicolumn{5}{c} {Total mass (dark matter+gas+galaxies)}\\
&&&&\\
\multicolumn{5}{c}{NFW profile ($\alpha$=1)} \\
Name &r$_{s}$& c$_{200}$ & R$_{200}$ &  M$_{200}$ & $\chi^{2}/$d.o.f.\\
     &Mpc& & Mpc & 10$^{14}$ M$_{\odot}$ &\\
&&&&\\
\hline
&&&&\\
A85 & 0.76$_{-0.17}^{+0.15}$&  3.55$_{-0.22}^{+0.26}$ & 2.71$_{-0.73}^{+0.65}$& 25.68$_{-3.18}^{+4.89}$&2.31\\
A119 &0.64$_{-0.22}^{+0.23}$& 4.12$_{-0.23}^{+0.25}$ & 2.66$_{-0.15}^{+0.13}$& 24.06$_{-2.18}^{+4.16}$ &1.57\\
A133 &0.29$_{-0.05}^{+0.15}$&8.11$_{-0.26}^{+0.25}$ & 2.33$_{-0.12}^{+0.19}$ & 16.34$_{-2.17}^{+3.55}$ &1.37\\
A168 &0.24$_{-0.10}^{+0.12}$&7.37$_{-0.27}^{+0.26}$ & 1.74$_{-0.56}^{+0.45}$ & 6.74$_{-1.78}^{+2.03}$ &2.11\\
A209 &0.65$_{-0.21}^{+0.18}$&3.11$_{-0.67}^{+0.77}$ & 2.01$_{-0.67}^{+0.54}$ & 12.05$_{-2.11}^{+2.34}$&1.03\\
A262 &0.13$_{-0.06}^{+0.07}$&10.11$_{-0.98}^{+1.02}$ & 1.33$_{-0.09}^{+0.12}$ & 2.93$_{-0.67}^{+0.73}$ &1.35\\
A383 &0.46$_{-0.17}^{+0.16}$&3.44$_{-0.36}^{+0.47}$ & 1.57$_{-0.17}^{+0.21}$ & 5.63$_{-1.19}^{+2.04}$ &2.02\\
A399 &1.43$_{-0.36}^{+0.28}$&2.15$_{-0.33}^{+0.37}$ & 3.07$_{-1.36}^{+1.37}$ &37.90$_{-4.11}^{+3.94}$ &2.47\\
A401 & 0.96$_{-0.37}^{+0.36}$& 3.19$_{-0.33}^{+0.38}$ & 3.07$_{-0.99}^{+1.21}$ &37.97$_{-2.81}^{+4.16}$&2.47\\
A478 &0.58$_{-0.21}^{+0.26}$&4.51$_{-0.48}^{+0.51}$ & 2.60$_{-0.55}^{+0.41}$ & 23.35$_{-2.17}^{+2.62}$ &0.84\\
A496 &0.17$_{-0.06}^{+0.09}$&11.26$_{-0.81}^{+0.83}$ & 1.93$_{-0.35}^{+0.48}$ & 9.10$_{-1.27}^{+1.19}$ &0.92\\
A521 &0.15$_{-0.04}^{+0.03}$&11.26$_{-0.78}^{+0.83}$ &1.67$_{-0.12}^{+0.16}$ & 7.21$_{-1.13}^{+1.16}$ &1.29\\
A539 &0.18$_{-0.04}^{+0.05}$&  10.37$_{-2.01}^{+2.04}$& 1.88$_{-0.18}^{+0.33}$& 8.38$_{-1.77}^{+1.46}$&3.32\\
A576 &0.59$_{-0.17}^{+0.16}$&  4.28$_{-0.81}^{+0.83}$&2.55$_{-0.73}^{+0.56}$ &  21.11$_{-1.19}^{+2.16}$&1.03\\
A644 &1.33$_{-0.29}^{+0.43}$&2.19$_{-0.27}^{+0.25}$ & 2.92$_{-0.81}^{+0.77}$& 32.55$_{-2.99}^{+3.51}$ &1.05\\
A697 &0.36$_{-0.13}^{+0.12}$&5.58$_{-1.22}^{+1.11}$ & 2.04$_{-0.22}^{+0.18}$& 13.56$_{-1.14}^{+1.27}$ &4.32\\
A754 &3.08$_{-0.48}^{+0.39}$&1.02$_{-0.22}^{+0.27}$ & 3.15$_{-1.15}^{+1.26}$ &  40.30$_{-3.16}^{+3.71}$&1.38\\
A773 &0.63$_{-0.13}^{+0.11}$&3.33$_{-0.53}^{+0.61}$ & 2.11$_{-0.35}^{+0.26}$ & 14.09$_{-1.28}^{+1.72}$&1.77\\
A907 &0.27$_{-0.08}^{+0.06}$&  6.25$_{-0.89}^{+1.01}$ & 1.67$_{-0.21}^{+0.15}$& 6.54$_{-0.72}^{+0.73}$&1.38\\
A963 & 0.41$_{-0.13}^{+0.15}$& 4.35$_{-0.99}^{+1.01}$ & 1.77$_{-0.18}^{+0.14}$& 8.22$_{-0.91}^{+1.24}$&2.83\\
A1060 &0.19$_{-0.03}^{+0.07}$&10.27$_{-2.17}^{+2.45}$ & 1.99$_{-0.34}^{+0.36}$& 9.81$_{-1.16}^{+1.32}$&2.16\\
A1068 &0.58$_{-0.19}^{+0.16}$&3.05$_{-0.22}^{+0.21}$ & 1.77$_{-0.06}^{+0.14}$ & 7.71$_{-0.78}^{+0.71}$&1.04\\
A1201 &0.25$_{-0.10}^{+0.11}$&6.27$_{-0.81}^{+0.83}$ & 1.56$_{-0.48}^{+0.51}$ & 5.43$_{-0.66}^{+0.59}$ &1.22\\
A1240&0.27$_{-0.10}^{+0.08}$&6.38$_{-0.15}^{+0.28}$ & 1.73$_{-0.58}^{+0.66}$ & 7.34$_{-0.81}^{+0.94}$ &1.11\\
A1361&0.28$_{-0.08}^{+0.04}$&7.28$_{-0.81}^{+0.83}$ & 2.01$_{-0.81}^{+0.84}$ & 11.07$_{-1.18}^{+1.34}$&1.00\\
A1413 &0.31$_{-0.10}^{+0.14}$&5.85$_{-0.44}^{+0.67}$& 1.83$_{-0.57}^{+0.66}$ & 8.53$_{-0.92}^{+1.18}$ &1.99\\
A1446&0.23$_{-0.03}^{+0.05}$&9.16$_{-0.72}^{-0.67}$ &2.11$_{-0.43}^{+0.48}$ &  12.66$_{-1.15}^{+1.26}$ &1.37\\
A1569 &0.21$_{-0.04}^{+0.05}$&10.37$_{-2.04}^{+2.01}$&2.22$_{-0.81}^{+0.84}$ & 14.35$_{-1.11}^{+1.34}$ &1.12\\
A1644&0.15$_{-0.08}^{+0.04}$&15.35$_{-4.17}^{+3.89}$& 2.25$_{-0.36}^{+0.39}$ & 14.60$_{-1.17}^{+1.47}$ &4.11\\
A1650&1.52$_{-0.29}^{+0.33}$&2.03$_{-0.37}^{+0.35}$ &3.10$_{-0.73}^{+0.72}$ &  39.44$_{-2.78}^{+3.74}$ &1.22\\
A1651 &0.57$_{-0.17}^{+0.16}$&4.29$_{-0.71}^{+0.74}$&2.45$_{-0.37}^{+0.46}$ & 19.48$_{-2.17}^{+2.45}$ &1.28\\
A1664 &0.35$_{-0.14}^{+0.11}$&7.24$_{-0.81}^{+0.83}$&2.55$_{-0.77}^{+0.72}$ & 22.84$_{-2.19}^{+2.81}$ &1.66\\
A1689 &0.27$_{-0.05}^{+0.05}$&8.35$_{-0.85}^{+0.91}$&2.23$_{-0.07}^{+0.03}$ & 16.09$_{-1.73}^{+2.81}$ &1.54\\
A1736 &0.11$_{-0.01}^{+0.00}$&16.27$_{-4.13}^{+4.16}$&1.85$_{-0.37}^{+0.42}$ & 8.11$_{-1.18}^{+1.15}$ &4.33\\
A1795 &0.53$_{-0.21}^{+0.20}$&4.61$_{-0.88}^{+0.55}$&2.46$_{-0.54}^{+0.36}$ & 19.34$_{-2.16}^{+2.18}$ &1.33\\
A1835 &0.92$_{-0.27}^{+0.18}$&2.66$_{-0.44}^{+0.35}$ &2.45$_{-0.24}^{+0.31}$ & 22.86$_{-3.11}^{+2.86}$ &2.77\\
A1914 &3.62$_{-0.37}^{+0.48}$&1.04$_{-0.33}^{+0.36}$&3.77$_{-0.67}^{+0.76}$ & 76.85$_{-5.28}^{+5.81}$ &3.28\\
A1995 &0.95$_{-0.28}^{+0.33}$&3.17$_{-0.33}^{+0.38}$ &3.01$_{-0.42}^{+0.43}$ & 45.32$_{-3.18}^{+5.16}$ &2.22\\
A2029 &0.42$_{-0.19}^{+0.17}$&6.99$_{-0.43}^{+0.41}$ &2.94$_{-0.82}^{+0.96}$ & 33.43$_{-4.12}^{+3.71}$ &1.20\\
A2034 &0.99$_{-0.15}^{+0.17}$&2.46$_{-0.62}^{+0.59}$ &2.45$_{-0.50}^{+0.51}$ & 19.98$_{-1.37}^{+2.54}$ &1.38\\
A2052 &0.19$_{-0.04}^{+0.03}$&10.33$_{-1.33}^{+1.39}$&1.93$_{-0.61}^{+0.55}$ & 9.12$_{-1.72}^{+1.26}$ &1.32\\
A2063 &0.27$_{-0.10}^{+0.14}$&7.36$_{-0.33}^{+0.38}$&2.01$_{-0.37}^{+0.45}$ &  10.30$_{-1.29}^{+1.94}$ &1.92\\
A2065 &1.31$_{0.18}^{+0.21}$&2.38$_{-0.56}^{+0.64}$&3.11$_{-0.37}^{+0.46}$ & 39.43$_{-4.18}^{+5.82}$ &1.20\\
A2124&0.19$_{-0.02}^{+0.04}$&11.36$_{-2.81}^{+2.84}$ &2.16$_{-0.61}^{+0.63}$ & 13.12$_{-1.27}^{+1.37}$ &1.47\\
A2142&0.54$_{-0.10}^{+0.13}$&5.27$_{-0.79}^{+0.83}$ & 2.82$_{-0.61}^{+0.59}$ & 29.87$_{-3.17}^{+2.47}$ &2.29\\
A2147&0.20$_{-0.03}^{+0.02}$&10.46$_{-4.27}^{+3.81}$ &2.05$_{-0.17}^{+0.14}$ & 10.93$_{-1.28}^{+1.27}$ &1.28\\
A2163&2.07$_{-0.35}^{+0.37}$&2.04$_{-0.33}^{+0.37}$&4.22$_{-0.66}^{+0.68}$ &111.18$_{-7.38}^{+5.92}$&1.75\\
A2199 &0.34$_{-0.10}^{+0.09}$&  6.27$_{-0.26}^{+0.25}$ & 2.14$_{-0.19}^{+0.28}$ &12.38$_{-1.29}^{+2.18}$&1.17\\
A2204 &0.90$_{-0.07}^{+0.11}$&  2.83$_{-0.44}^{+0.32}$ & 2.55$_{-0.66}^{+0.71}$ &23.35$_{-2.17}^{+3.16}$&1.28\\
A2218 &0.26$_{-0.03}^{+0.01}$&  6.33$_{-1.55}^{+2.34}$ & 1.65$_{-0.12}^{+0.16}$ & 6.46$_{-0.83}^{+0.82}$&1.10\\
A2219 &0.84$_{-0.17}^{+0.15}$&  3.44$_{-0.73}^{+0.73}$ & 2.89$_{-0.91}^{+0.78}$ &36.49$_{-3.39}^{+4.92}$&2.18\\
A2244 &1.24$_{-0.17}^{+0.15}$&  2.43$_{-0.81}^{+0.84}$ & 3.01$_{-0.39}^{+0.51}$ &36.52$_{-3.18}^{+3.16}$&4.19\\
A2256 &2.87$_{-0.37}^{+0.44}$&  1.16$_{-0.27}^{+0.37}$ & 3.33$_{-0.58}^{+0.62}$ &47.78$_{-5.29}^{+5.37}$&2.11\\
&&&&\\
\hline
\end{tabular}
\end{table*}

\setcounter{table}{1}
\begin{table*}
\centering
\fontsize{7}{7}
\selectfont
\caption{\fontsize{9}{9} \selectfont Continued.}
\bigskip
\begin{tabular}{lccccc}
\hline
&&&&\\
\multicolumn{5}{c} {Total mass (dark matter+gas+galaxies)}\\
&&&&\\
\multicolumn{5}{c}{NFW profile ($\alpha$=1)} \\
Name &r$_{s}$& c$_{200}$ & R$_{200}$ &   M$_{200}$ & $\chi^{2}/$d.o.f.\\
     &Mpc& & Mpc & 10$^{14}$ M$_{\odot}$ &\\
&&&&\\
\hline
&&&&\\
A2319 &2.57$_{-0.19}^{+0.32}$&  1.28$_{-0.22}^{+0.27}$ & 3.29$_{-1.29}^{+1.12}$ &46.00$_{-5.82}^{+4.57}$&1.22\\
A2390 &1.26$_{-0.16}^{+0.14}$&  2.11$_{-0.62}^{+0.52}$ & 2.65$_{-0.54}^{+0.61}$ &28.21$_{-2.39}^{+3.15}$&1.02\\
A2462 &0.19$_{-0.06}^{+0.03}$&  13.23$_{-0.67}^{+0.72}$ & 2.51$_{-0.22}^{+0.27}$&20.73$_{-2.49}^{+2.47}$&1.05\\
A2537 &0.44$_{-0.14}^{+0.12}$&  4.88$_{-0.24}^{+0.24}$ & 2.15$_{-0.55}^{+0.63}$&16.12$_{-2.19}^{+2.16}$ &1.66\\
A2589 &0.32$_{-0.06}^{+0.09}$&  6.27$_{-0.72}^{+0.75}$ & 1.99$_{-0.51}^{+0.63}$ &10.05$_{-1.29}^{+1.27}$&1.48\\
A2597 &0.24$_{-0.03}^{+0.05}$&  8.28$_{-1.26}^{+1.27}$ & 2.01$_{-0.73}^{+0.57}$ &10.76$_{-1.22}^{+1.27}$&2.12\\
A2634 &0.18$_{-0.03}^{+0.05}$&  11.38$_{-2.16}^{+2.17}$ & 2.11$_{-0.11}^{+0.05}$&11.88$_{-1.27}^{+2.15}$&1.47\\
A2657 &0.36$_{-0.10}^{+0.11}$&  5.49$_{-0.81}^{+0.83}$ & 1.98$_{-0.11}^{+0.15}$&  9.89$_{-1.28}^{+2.17}$&1.55\\
A2667 &1.04$_{-0.22}^{+0.23}$&  2.25$_{-0.15}^{+0.17}$ & 2.35$_{-0.36}^{+0.45}$ & 19.71$_{-1.27}^{+2.61}$&1.28\\
A2670 &0.44$_{-0.16}^{+0.13}$&  5.33$_{-1.09}^{+1.11}$ & 2.33$_{-1.02}^{+1.03}$ & 16.62$_{-1.62}^{+1.25}$&2.21\\
A2717 &0.35$_{-0.04}^{+0.05}$&  6.12$_{-0.81}^{+0.84}$ & 2.16$_{-0.81}^{+0.79}$ & 12.94$_{-1.26}^{+2.11}$&1.83\\
A2744 &1.38$_{-0.27}^{_0.30}$&  1.72$_{-0.17}^{+0.15}$ & 2.38$_{-0.31}^{+0.36}$ & 22.17$_{-2.26}^{+2.14}$&1.36\\
A3112 &0.26$_{-0.03}^{+0.04}$&  9.36$_{-0.73}^{+0.73}$ & 2.43$_{-0.26}^{+0.31}$ & 18.84$_{-2.16}^{+3.14}$&3.27\\
A3158 &1.24$_{-0.14}^{+0.13}$&  2.19$_{-0.81}^{+0.80}$ & 2.71$_{-0.62}^{+0.38}$ & 25.80$_{-2.72}^{+3.17}$&3.22\\
A3266 &0.35$_{-0.05}^{+0.05}$&  8.36$_{-0.81}^{+0.83}$ & 2.95$_{-0.77}^{+0.52}$ & 33.25$_{-3.27}^{+4.16}$&3.21\\
A3376 &0.37$_{-0.04}^{+0.03}$&  7.38$_{-1.03}^{+1.01}$ & 2.75$_{-0.62}^{+0.43}$ & 26.61$_{-2.91}^{+3.72}$&1.29\\
A3391 &0.26$_{-0.03}^{+0.05}$&  9.26$_{-1.36}^{+1.14}$ & 2.45$_{-0.35}^{+0.84}$ & 18.91$_{-1.28}^{+1.25}$&1.02\\
A3395 &0.79$_{-0.10}^{+0.11}$&  3.74$_{-0.37}^{+0.35}$ & 2.97$_{-0.84}^{+0.96}$ & 33.67$_{-4.12}^{+4.27}$&1.07\\
A3526 &0.18$_{-0.03}^{+0.04}$&  10.43$_{-0.83}^{+0.81}$ & 1.88$_{-0.27}^{+0.25}$ & 8.26$_{-0.28}^{+0.72}$&1.66\\
A3558 &0.30$_{-0.07}^{+0.04}$&  8.37$_{-0.83}^{+0.78}$ & 2.55$_{-0.36}^{+0.38}$ & 21.27$_{-2.16}^{+3.27}$&2.33\\
A3562 &0.27$_{-0.06}^{+0.05}$&  8.26$_{-0.27}^{+0.25}$ & 2.22$_{-0.17}^{+0.15}$ & 14.05$_{-1.46}^{+1.56}$&2.17\\
A3571 &0.32$_{-0.03}^{+0.06}$&  8.92$_{-2.10}^{+1.98}$ & 2.88$_{-0.37}^{+0.46}$ & 30.41$_{-3.84}^{+3.75}$&1.36\\
A3667 &0.29$_{-0.04}^{+0.06}$&  9.36$_{-1.36}^{+1.25}$ & 2.75$_{-0.52}^{+0.61}$ &  26.84$_{-2.46}^{+2.48}$&1.22\\
A3827 &0.87$_{-0.13}^{+0.09}$&  4.45$_{-0.77}^{+0.67}$ & 3.89$_{-0.43}^{+0.16}$ &  78.90$_{-4.27}^{+5.26}$&1.02\\
A3921 &0.53$_{-0.07}^{+0.08}$&   5.27$_{-0.71}^{+0.73}$ & 2.81$_{-0.26}^{+0.33}$ & 29.61$_{-1.77}^{+2.15}$&1.33\\
A4038 &0.20$_{-0.01}^{+0.02}$&  9.38$_{-1.01}^{+1.03}$ & 1.87$_{-0.17}^{+0.25}$ & 8.26$_{-0.92}^{+1.18}$&1.32\\
A4059 &0.22$_{-0.00}^{+0.02}$&  9.36$_{-2.01}^{+2.02}$ & 2.03$_{-0.34}^{+0.28}$ & 10.72$_{-1.28}^{+0.92}$&1.37\\
AWM4 & 0.24$_{-0.02}^{+0.03}$& 7.21$_{-1.03}^{+1.02}$ & 1.73$_{-0.22}^{+0.27}$ &  6.55$_{-0.37}^{+0.85}$&1.22\\
CLJ1226.9+3332 &1.75$_{-0.11}^{+0.13}$&  2.04$_{-0.36}^{+0.32}$ & 3.57$_{-0.52}^{+0.51}$ & 143.19$_{-17.26}^{+15.26}$&1.45\\
Coma (A1656) &2.35$_{-0.55}^{+0.43}$&  1.37$_{-0.27}^{+0.25}$ & 3.22$_{-0.62}^{+0.59}$ & 41.94$_{-3.92}^{+4.72}$&1.44\\
IIIZw54 &0.17$_{-0.02}^{+0.01}$ & 11.25$_{-1.38}^{+1.33}$ & 1.91$_{-0.43}^{+0.51}$ &  8.81$_{-0.93}^{+1.04}$&1.66\\
ISCS J1438.1+3414 &6.81$_{-0.72}^{+0.55}$&  0.55$_{-0.08}^{+0.05}$ & 3.75$_{-0.38}^{+0.33}$ & 293.15$_{-36.84}^{+18.27}$ &1.73\\
MACSJ0011.7-1523 & 0.75$_{-0.23}^{+0.13}$&4.01$_{-0.33}^{+0.23}$ & 3.01$_{-0.31}^{+0.30}$ &48.34$_{-4.15}^{+3.82}$ &1.43\\
MACSJ0159.8-0849 & 0.54$_{-0.13}^{+0.12}$&5.35$_{-0.72}^{+0.82}$ & 2.89$_{-0.52}^{+0.51}$ &44.00$_{+4.44}^{+4.84}$ &1.49\\
MACSJ0242.6-2132 & 0.32$_{-0.04}^{+0.06}$& 7.88$_{-1.23}^{+1.64}$ & 2.55$_{-0.45}^{+0.46}$ &27.44$_{-2.19}^{+1.99}$ &1.33\\
MACSJ0429.6-0253 & 0.65$_{-0.10}^{+0.13}$&3.36$_{-1.65}^{+1.76}$ & 2.17$_{-0.84}^{+0.78}$ &18.50$_{-1.63}^{+1.82}$ &1.38\\
MACSJ0647.7+7015 & 3.42$_{-0.36}^{+0.35}$& 1.00$_{-0.11}^{+0.12}$ & 3.42$_{-0.54}^{+0.45}$ & 89.51$_{-9.28}^{+10.14}$ &1.22\\
MACSJ0744.8+3927 &  3.44$_{-0.37}^{+0.39}$&1.01$_{-0.16}^{+0.37}$ & 3.48$_{-0.47}^{+0.45}$ & 106.48$_{-6.18}^{+8.82}$ &1.28\\
MACSJ1115.8+0129 & 1.07$_{-0.11}^{+0.09}$& 2.23$_{-0.55}^{+0.53}$ & 2.39$_{-0.63}^{+0.64}$ &23.59$_{-3.18}^{+4.17}$ &3.21\\
MACSJ1311.0-0311 & 0.60$_{-0.06}^{+0.07}$& 5.01$_{-0.35}^{+0.73}$ & 3.02$_{-0.35}^{+0.33}$ & 55.11$_{-5.58}^{+4.84}$ &2.18\\
MACSJ1423.8+2404 & 0.85$_{-0.22}^{+0.29}$& 3.33$_{-0.63}^{+0.65}$ & 2.83$_{-0.32}^{+0.17}$ & 48.21$_{-4.72}^{+4.72}$ &1.88\\
MACSJ1427.6-2521 & 0.30$_{-0.32}^{+0.17}$& 8.28$_{-1.77}^{+1.99}$ & 2.47$_{-0.84}^{+0.81}$ &25.04$_{-2.26}^{+2.36}$ &2.18\\
MACSJ1720.3+3536 & 0.54$_{-0.11}^{+0.10}$& 5.25$_{-0.36}^{+0.61}$ & 2.83$_{-0.67}^{+0.68}$ & 40.70$_{-4.92}^{+5.17}$& 1.72\\
MACSJ1931.8-2635 & 0.65$_{-0.06}^{+0.05}$& 4.11$_{-0.24}^{+0.36}$ & 2.67$_{-0.92}^{+0.91}$ &32.78$_{-4.19}^{+3.91}$ &1.44\\ 
MACSJ2129.4-0741 & 2.06$_{-0.27}^{+0.32}$& 1.70$_{-0.41}^{+0.38}$ & 3.51$_{-0.72}^{+0.71}$ & 96.66$_{-10.15}^{+10.26}$&1.46\\
MACSJ2229.8-2756 &0.35$_{-0.03}^{+0.06}$ & 8.54$_{-1.36}^{+1.67}$ & 3.02$_{-0.84}^{+0.81}$ &46.06$_{-5.17}^{+4.92}$ &1.33\\
MKW3s & 0.18$_{-0.03}^{+0.03}$ &11.37$_{-2.73}^{+2.18}$ & 2.11$_{-0.18}^{+0.12}$ & 12.02$_{-3.19}^{+2.47}$&1.22\\
MKW4 & 0.06$_{-0.00}^{+0.01}$& 19.93$_{-3.17}^{+3.15}$ & 1.21$_{-0.04}^{+0.10}$ &  2.22$_{-2.10}^{+1.84}$&1.37\\
MKW8 & 0.12$_{-0.01}^{+0.02}$& 15.28$_{-3.16}^{+3.14}$ & 1.81$_{-0.28}^{+0.33}$ &  7.47$_{-2.57}^{+1.48}$&1.92\\
PKS0745-191 &0.45$_{-0.06}^{+0.06}$& 6.45$_{-0.61}^{+0.73}$&2.92$_{-0.18}^{+0.22}$& 33.52$_{-3.17}^{+3.17}$&1.82\\
RXCJ0043.4-2037 &0.19$_{-0.01}^{+0.03}$&  8.01$_{-3.03}^{+3.89}$ & 1.58$_{-0.17}^{+0.17}$ & 6.38$_{-0.84}^{+0.82}$ &1.11\\
RXCJ0232.2-4420 &1.19$_{-0.16}^{+0.13}$&  1.88$_{-0.66}^{+0.67}$ & 2.25$_{-0.43}^{+0.34}$ & 18.25$_{-1.82}^{+2.16}$ &1.22\\
RXCJ0307.0-2840 &0.62$_{-0.06}^{+0.07}$&  3.22$_{-0.77}^{+0.88}$ & 2.01$_{-0.29}^{+0.26}$ & 12.62$_{-1.82}^{+1.72}$ &1.38\\
RXCJ0516.7-5430 &0.83$_{-0.06}^{+0.09}$&  2.45$_{-0.77}^{+1.55}$ & 2.05$_{-0.25}^{+0.32}$ & 13.96$_{-1.22}^{+1.54}$ &1.72\\
RXCJ0547.6-3152 &0.48$_{-0.07}^{+0.06}$&  4.14$_{-0.63}^{+0.57}$ & 1.99$_{-0.43}^{+0.14}$ & 11.06$_{-1.10}^{+1.17}$ &1.34\\
RXCJ0605.8-3518 &0.40$_{-0.01}^{+0.02}$&  4.14$_{-0.23}^{+0.33}$ & 1.66$_{-0.15}^{+0.18}$ & 6.37$_{-0.72}^{+0.75}$ &1.38\\
RXCJ1131.9-1955 &0.83$_{-0.02}^{+0.11}$&  2.55$_{-1.67}^{+1.24}$ & 2.11$_{-0.28}^{+0.34}$ & 15.43$_{-1.66}^{+1.45}$ &1.88\\

&&&&\\
\hline 
\end{tabular}
\end{table*}

\setcounter{table}{1}
\begin{table*}
\centering
\fontsize{7}{7}
\selectfont
\caption{\fontsize{9}{9} \selectfont Continued.}
\bigskip
\begin{tabular}{lcccccc}
\hline
&&&&\\
\multicolumn{5}{c} {Total mass (dark matter+gas+galaxies)}\\
&&&&\\
\multicolumn{5}{c}{NFW profile ($\alpha$=1)} \\
Name &r$_{s}$& c$_{200}$ & R$_{200}$ & M$_{200}$ & $\chi^{2}/$d.o.f.\\
     &Mpc& & Mpc & 10$^{14}$ M$_{\odot}$ &\\
&&&&\\
\hline
&&&&\\
RXCJ2014.8-2430 &0.50$_{-0.04}^{+0.07}$&  3.88$_{-0.67}^{+0.54}$ & 1.95$_{-0.17}^{+0.12}$ & 10.53$_{-1.47}^{+1.48}$ &2.83\\
RXCJ2129.6+0005 &0.47$_{-0.06}^{+0.07}$&  3.77$_{-0.33}^{+0.38}$ & 1.77$_{-0.21}^{+0.22}$ & 8.46$_{-1.10}^{+0.94}$ &2.33\\
RXCJ2337.6+0016 &0.35$_{-0.03}^{+0.06}$&  5.02$_{-1.77}^{+1.55}$ & 1.75$_{-0.18}^{+0.22}$ & 8.50$_{-0.89}^{+1.01}$ &2.18\\
ZwCL1215 &0.52$_{-0.06}^{+0.08}$&  5.62$_{-0.40}^{+0.42}$ & 2.95$_{-0.63}^{+0.56}$ & 33.71$_{-4.29}^{+4.17}$&1.04\\
RCSJ0224-0002 &0.56$_{-0.07}^{+0.09}$&2.78$_{-0.25}^{+0.25}$&1.56$_{-0.17}^{+0.16}$&10.51$_{-1.28}^{+1.11}$&1.09\\
RCSJ0439-2904&0.25$_{-0.03}^{+0.01}$ &4.12$_{-0.21}^{+0.25}$&1.04$_{-0.12}^{+0.14}$&3.79$_{-0.27}^{+0.56}$&1.20\\
RCSJ1107-0523&0.40$_{-0.03}^{+0.03}$ &3.15$_{-0.45}^{+0.56}$&1.26$_{-0.27}^{+0.29}$&5.27$_{-0.45}^{+0.74}$&1.05\\
RCSJ1419.2+5326& 0.28$_{-0.05}^{+0.03}$&  6.24$_{-0.71}^{+0.73}$&1.76$_{-0.21}^{+0.28}$& 12.91$_{-1.72}^{+1.64}$ &1.48\\
RCSJ1620+2929 &0.31$_{-0.06}^{+0.05}$&4.38$_{-0.71}^{+0.72}$&1.35$_{-0.17}^{+0.18}$&7.57$_{-1.04}^{+1.02}$&1.47\\
RCS2156+0123&1.93$_{-0.24}^{+0.12}$ &1.23$_{-0.30}^{+0.41}$&2.38$_{-0.29}^{+0.28}$&22.80$_{-2.88}^{+3.28}$&1.05\\
RCSJ2318+0034&1.83$_{-0.15}^{+0.13}$ &1.18$_{-0.17}^{+0.19}$&2.17$_{-0.31}^{+0.37}$&28.37$_{-3.18}^{+4.11}$&2.10\\
RCSJ2319+0038 &0.71$_{-0.09}^{+0.11}$&3.27$_{-0.55}^{+0.56}$&2.34$_{-0.29}^{+0.48}$&40.96$_{-5.22}^{+4.25}$&1.36\\
RXJ0439.0+0520& 0.24$_{-0.03}^{+0.05}$& 7.71$_{-1.77}^{+1.27}$&1.83$_{-0.36}^{+0.48}$&9.11$_{-2.27}^{+1.75}$&1.46\\
RXJ0848.7+4456 &1.40$_{-0.17}^{+0.14}$& 0.82$_{-0.01}^{+0.05}$&1.15$_{-0.13}^{+0.15}$& 3.32$_{-0.77}^{+0.73}$&2.66\\
RXJ0849+4452 &1.89$_{-0.25}^{+0.27}$ & 1.14$_{-0.10}^{+0.11}$&2.16$_{-0.51}^{+0.52}$&4.67$_{-0.47}^{+0.66}$&1.88\\
RXJ0910+5422 &0.96$_{-0.09}^{+0.15}$& 2.64$_{-0.21}^{+0.25}$&2.53$_{-0.63}^{+0.63}$&64.92$_{-7.34}^{+5.67}$&1.46\\
RXJ1113.1-2615 &0.38$_{-0.04}^{+0.06}$&  3.28$_{-0.71}^{+0.73}$&1.26$_{-0.11}^{+0.15}$&5.30$_{-0.48}^{+0.73}$& 1.22\\
RXJ1221.4+4918&1.10$_{-0.10}^{+0.15}$&2.39$_{-0.35}^{+0.37}$&2.64$_{-0.35}^{+0.31}$&46.64$_{-3.49}^{+5.22}$&1.07\\
&&&&&\\
\hline 
\end{tabular}
\end{table*}


As in SA07, we define the concentration parameters as $c_{vir} = (r^{total}_{vir}/r^{dark}_{s})$\footnote{Both virial radii and virial masses are calculated for the total mass model, including all mass components.}. 

\begin{figure}
\centering
\epsfig{file=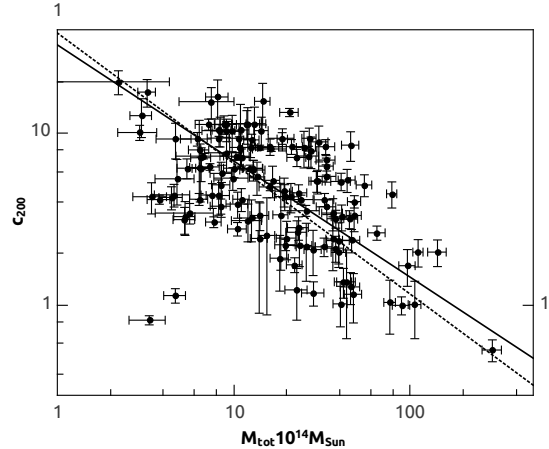, width=0.9\linewidth}
\caption{ The mass-concentration relation for the total mass (c$_{200}$-M$_{200}$). 
Solid line: best-fitting model with $c_{0} = $ 6.83 $ \pm$ 0.64 and $a = $ -0.19 $ \pm$ 0.07 ($b$ = 1 fixed) for model \ref{fit1}. 
Dashed line: best-fitting model with   $c_{0} = $ 7.73 $ \pm$ 1.07, $a = $ -0.56 $ \pm $ 0.15 and $b = $ 0.80 $ \pm $ 0.25 for model \ref{fit2} (both have 68 per cent confidence level).}
\label{cm200}
\end{figure}

The mass-concentration relation measured from the Chandra, shown in Fig. \ref{cm200}, 
shows that the mass-concentration relation decreases when mass increases.
The simplest analytic form that describe the mass-concentration relation is a power-law model (\citealt{Dolag:04})
\begin{equation}
\label{fit1}
c(z) = \frac{c_{0}}{1+z}(\frac{M}{8 \times 10^{14} h^{-1}M_{\odot}})^{a}.
\end{equation}

For example, \citet{Shaw:06} by using the previous relation, with their simulated data, found $c_{0}$ = 6.47$ \pm$ 0.03 and $a$ = -0.12$ \pm$ 0.03 for 68 per cent confidence limits in agreement with \citet{Eke:01}, and  \citet{Bullock:01}. 
In a recent work by Kulinich et al. (2012), a new method to calculate a halo concentration parameter, and taking into account the halo overdensity and merging, was proposed. 
In our research, we used also a power-law model, obtaining $c_{0} = $ 6.83$ \pm$ 0.64 and $a = $ -0.19$ \pm$ 0.07 ($b$ = 1 fixed). 
As we show in Fig. \ref{cm200}, using the measurements obtained with our method above, the relation between concentration and total masses for CDM halos is represented not so well from Dolag et al. (2004) formula (Eq. 22). In agreement with SA07, Fig. 1, 
the model with $b=1$, provides a poor fit for the observations, lying systematically below the data at lower masses and above it in the highest mass range. 

In order to improve the description of Chandra data, we assumed, as SA07, differently from \citet{Bullock:01} to model
a redshift evolution of the form $(1+z)^{-b}$, with $b$ free 
\begin{equation}
\label{fit2}
c = \frac{c_{0}}{(1+z)^{b}}\left( \frac{M_{200}}{8 \times 10^{14} h^{-1}M_{\odot}}\right)^{a}.
\label{eq:cc}
\end{equation}
The motivation to the new redshift dependence comes from the fact that clusters are more complicated objects than those assumed in \citet{Bullock:01} (see discussion in Sec. 3.2 of Sa07, and Zhao et al. 2003).
%

The results from a fit using Eq. (\ref{eq:cc}), with $c_{0}, a$ and $b$ all free, are shown by the dashed line in Fig. \ref{cm200}. This model provides an improved description of the data with $\chi^{2} = $1.3 and best-fitting parameteres $c_{0} = $7.73$ \pm$ 1.07, $a = $ -0.56 $\pm$ 0.15 and $b = $0.80$ \pm$ 0.25 (68 per cent confidence limits). Note, that RXJ0848.7+4456 and RXJ0849+4452 have been excluded from the fitting, since they give small values for concentration parameter ($c_{200}$,  0.82$_{-0.01}^{+0.05}$ and 1.14$_{-0.10}^{+0.11}$, respectively).
\begin{figure}
\centering
\epsfig{file=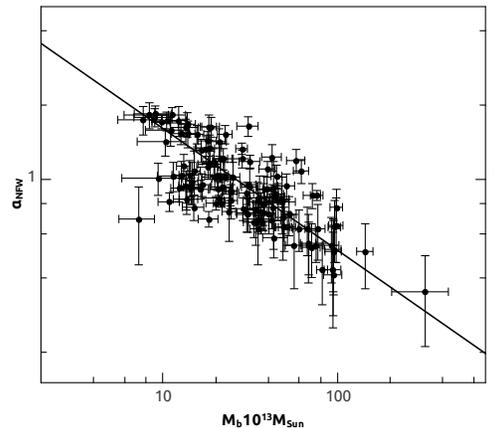, width=0.8\linewidth}
\caption{$\alpha$-M$_{b}$ relation for all sample of galaxy clusters. The data were obtained fitting the DM profiles of clusters with the gNFW model. The baryonic mass was obtained, as described in the text, by subtracting the DM component from the total mass.}
\label{alphamb}
\end{figure}

\subsection{The inner slope of DM profiles}

Finally, we fitted the DM profiles by means of a gNFW model to have hints on the inner slopes of the density profiles. 
The results are summarized in Tab. \ref{res4}. 
The value of $\alpha$ obtained is $\alpha $ = 0.94 $\pm$ 0.13 (68 per cent confidence limits). This result is consistent with CDM predictions.

The values of the slope are in the range $0.5 < \alpha < 1.8 $ and for the largest number of clusters the density profile is compatible with the expectation from $\Lambda$CDM model $0.7<\alpha<1.5$ (SA07). Only some clusters have slopes smaller than 0.7 and they are usually characterized by an higher $M_b$ with respect to clusters with larger $\alpha$. 

\subsection{Inner slope and baryons content}

As discussed in the introduction not all clusters have profiles best fitted by the NFW model (e.g., \citealt{Sand:02}, \citealt{Sand:04}, \citealt{Sand:08}; \citealt{Newman:09}, \citealt{Newman:11}). Among the processes able to produce flat profile, the role of baryons has been several times claimed (see the introduction). Recently, \citet{delPopolo:12} studied the effect of baryons, and the central BCG in shaping the inner density profile. It was also shown how higher content of baryons can give rise to profiles similar to A611 and A383 which have slopes flatter than NFW profile, and small baryonic content can produce steeper inner profiles in agreement with the NFW model (MACS J1423.8+2404, RXJ1133). 

Motivated by the previous arguments, we studied the eventual correlation among the inner density profile slope, $\alpha$, and the baryonic content of the clusters that we studied. 

Similarly to SA07, we performed an analysis in which the total mass was separated in the DM halo, the X-ray emitting halo, and the luminous mass of the BCG (this is the second case mentioned in Sect. 4.2). We fitted the DM profile with a gNFW model and calculated $\alpha$, and reported the values in Tab. \ref{res4}. From the analysis the baryonic content, $M_b$ is calculated summing the diffuse gas content and the baryons contained in the central cD galaxy in each cluster. In Fig. \ref{alphamb}, we plot the inner slope, $\alpha$, versus the baryonic content at the radius of $R_{200}$. A2255 has been excluded during fitting, because this galaxy cluster gave small uncertain value of $\chi^{2}$. The plot shows a tight correlation between the inner slope $\alpha$ and the baryonic content: the larger is the baryonic content the flatter is the inner profile. This results is in agreement with studies of the role of baryons on the inner slope of clusters (e.g., \citealt{delPopolo:09}, \citealt{delPopolo:12}, \citealt{Romano-Diaz:08}, \citealt{Romano-Diaz:09}).  

An important issue to stress is the fact that we assumed, as SA07 that all the central BCG has the same mass, so that the difference 
of the baryon content from cluster to cluster is connected to the diffused gas content. The expectation that the central baryonic content is the only or the fundamental contribution to shape the inner density profile is only partly true. To understand this, 
it is interesting to discuss more in detail the role of the total baryonic mass, $M_b$, and the central one. 

\begin{table*}
\centering
\fontsize{7}{7}
\selectfont
\caption{\fontsize{9}{9} \selectfont Results on dark matter profiles for the clusters, from the fits that includes DM, X-ray emitting gas, and BCG. 
The first column includes names of galaxy cluster, while the second, third and sixth column refers to the NFW fit: second shows the $r_{s}$ parameter from NFW model for dark matter, third shows the concentration parameter $c_{200}=R_{200}/R_{s}$, sixt the $\chi^{2}/$d.o.f. goodness of fit.
In the case of the gNFW, we give the inner value of $\alpha$, in seventh column, while in eighth column is reported 
the $\chi^{2}/$d.o.f. goodness of fit. Fourth column shows the mass of baryonic matter in clusters, and fifth 
shows the fraction of baryonic matter. 
Confidence limits are 68 per cent.}
\bigskip
\label{res4}
\begin{tabular}{lccccccc}
\hline
&&&&&&&\\
\multicolumn{8}{c} {Dark matter profiles}\\
&&&&&&&\\
\multicolumn{6}{c}{NFW profile ($\alpha$=1)} &\multicolumn{2}{c}{NFW ($\alpha$ is free parameter)}\\
Name &r$_{s}$& c$_{200}$& M$_{b}$& $f_{M_{b}/M_{tot}}$ & $\chi^{2}/$d.o.f. &$\alpha$&$\chi^{2}/$d.o.f.\\
     &Mpc& & 10$^{13} M_{\odot}$ & & & &\\
&&&&&&&\\
\hline
&&&&&&&\\
A85 & 0.69$_{-0.20}^{+0.13}$&  3.15$_{-0.32}^{+0.23}$&42.86$_{-6.46}^{+4.83}$ &0.166&1.11 &0.85$_{-0.28}^{+0.16}$&2.11\\
A119 &0.54$_{-0.12}^{+0.13}$& 3.83$_{-0.33}^{+0.35}$ &47.97$_{-9.12}^{+8.35}$&0.199&2.46&0.82$_{-0.25}^{+0.27}$&2.57\\
A133 &0.24$_{-0.06}^{+0.13}$&7.83$_{-0.46}^{+0.35}$ &35.02$_{-6.45}^{+3.87}$&0.214&1.55 &0.69$_{-0.08}^{+0.08}$&1.22\\
A168 &0.18$_{-0.06}^{+0.07}$&7.03$_{-0.20}^{+0.17}$ &19.76$_{-3.17}^{+2.17}$&0.293&1.37 &1.01$_{-0.11}^{+0.17}$&2.04\\
A209 &0.58$_{-0.11}^{+0.08}$&2.86$_{-0.37}^{+0.27}$ &10.47$_{-1.82}^{+1.85}$&0.087& 2.01&1.72$_{-0.13}^{+0.16}$&1.53\\
A262 &0.11$_{-0.02}^{+0.03}$&9.75$_{-0.93}^{+1.11}$ &7.25$_{-1.67}^{+1.53}$&0.246&1.64&0.69$_{-0.04}^{+0.03}$&1.39\\
A383 &0.43$_{-0.16}^{+0.13}$&3.32$_{-0.32}^{+0.27}$ &11.03$_{-2.18}^{+3.11}$&0.195&1.55 &1.58$_{-0.22}^{+0.22}$&1.02\\
A399 &1.22$_{-0.46}^{+0.38}$&2.01$_{-0.43}^{+0.47}$ &94.53$_{-9.16}^{+10.11}$&0.249&1.89 &0.51$_{-0.15}^{+0.17}$&2.47\\
A401 & 0.84$_{-0.27}^{+0.26}$& 3.00$_{-0.23}^{+0.28}$ &75.91$_{-5.83}^{+6.31}$&0.199&2.05&0.86$_{-0.37}^{+0.29}$&2.13\\
A478 &0.51$_{-0.11}^{+0.16}$&4.04$_{-0.58}^{+0.41}$ &30.86$_{-3.17}^{+2.88}$&0.132&1.44 &0.96$_{-0.16}^{+0.22}$&1.24\\
A496 &0.13$_{-0.06}^{+0.06}$&10.73$_{-1.81}^{+1.83}$ &10.86$_{-1.22}^{+1.51}$&0.119&2.46&0.82$_{-0.07}^{+0.05}$&1.42\\
A521 &0.11$_{-0.05}^{+0.05}$&10.77$_{-1.28}^{+1.43}$ &14.47$_{-2.17}^{+1.95}$&0.200& 1.42&0.93$_{-0.13}^{+0.12}$&1.19\\
A539 &0.15$_{-0.05}^{+0.04}$&  10.21$_{-1.21}^{+1.44}$&15.06$_{-1.19}^{+1.25}$&0.179&1.79&1.08$_{-0.22}^{+0.21}$&2.32\\
A576 &0.51$_{-0.12}^{+0.13}$&  4.03$_{-0.21}^{+0.33}$&23.87$_{-2.11}^{+2.14}$&0.113& 1.11&0.74$_{-0.18}^{+0.18}$&1.63\\
A644 &1.12$_{-0.19}^{+0.33}$&2.00$_{-0.21}^{+0.22}$ &44.62$_{-2.91}^{+3.17}$& 0.137& 1.37&1.03$_{-0.06}^{+0.11}$&1.55\\
A697 &0.20$_{-0.05}^{+0.10}$&5.21$_{-0.52}^{+0.61}$ &13.49$_{-1.19}^{+1.17}$&0.099& 0.99&0.94$_{-0.13}^{+0.14}$&4.32\\
A754 &2.84$_{-0.38}^{+0.29}$&0.84$_{-0.12}^{+0.17}$ &61.80$_{-5.19}^{+6.17}$& 0.153& 1.53&1.08$_{-0.09}^{+0.12}$&1.98\\
A773 &0.55$_{-0.10}^{+0.09}$&3.04$_{-0.33}^{+0.41}$ & 24.47$_{-1.98}^{+2.15}$&0.173& 1.73&0.82$_{-0.11}^{+0.13}$&4.22\\
A780 &0.15$_{-0.03}^{+0.05}$&  9.05$_{-0.73}^{+0.81}$ &40.35$_{-3.78}^{+3.71}$&0.228& 2.28&0.92$_{-0.07}^{+0.04}$&0.77\\
A907 &0.22$_{-0.05}^{+0.04}$&  6.01$_{-0.80}^{+1.11}$ &13.12$_{-1.19}^{+1.13}$&0.200& 2.00&1.12$_{-0.12}^{+0.13}$&1.18\\
A963 & 0.35$_{-0.11}^{+0.10}$& 4.13$_{-0.69}^{+0.51}$ &22.63$_{-2.19}^{+2.14}$&0.275& 2.75&1.52$_{-0.12}^{+0.13}$&1.83\\
A1060 &0.15$_{-0.02}^{+0.05}$&9.58$_{-1.17}^{+1.45}$ &24.28$_{-1.93}^{+2.16}$&0.247& 2.47&0.89$_{-0.11}^{+0.08}$&3.16\\
A1068 &0.53$_{-0.13}^{+0.11}$&2.75$_{-0.32}^{+0.51}$ &14.64$_{-1.19}^{+1.16}$&0.189&  1.89&0.98$_{-0.19}^{+0.19}$&1.54\\
A1201 &0.20$_{-0.06}^{+0.10}$&5.88$_{-0.61}^{+0.63}$ &9.79$_{-0.93}^{+1.15}$& 0.180& 1.80&1.72$_{-0.10}^{+0.10}$&2.22\\
A1240&0.21$_{-0.10}^{+0.08}$&6.11$_{-0.25}^{+0.18}$ &16.42$_{-1.67}^{+1.73}$&0.223&  2.23&0.91$_{-0.08}^{+0.08}$&2.21\\
A1361&0.22$_{-0.06}^{+0.06}$&6.81$_{-0.51}^{+0.63}$ & 21.56$_{-2.17}^{+2.15}$&0.194& 1.94&0.80$_{-0.28}^{+0.29}$&1.50\\
A1413 &0.24$_{-0.11}^{+0.10}$&5.85$_{-0.44}^{+0.67}$&14.48$_{-1.18}^{+1.42}$& 0.169&1.69&0.93$_{-0.08}^{+0.08}$&1.99\\
A1446&0.21$_{-0.03}^{+0.05}$&9.06$_{-0.72}^{-0.67}$ &28.20$_{-2.17}^{+2.16}$& 0.222&2.22&1.34$_{-0.72}^{+0.73}$&2.37\\
A1569 &0.19$_{-0.04}^{+0.05}$&10.37$_{-2.04}^{+2.01}$&37.01$_{-3.17}^{+3.26}$&0.257& 2.57&0.67$_{-0.19}^{+0.18}$&2.12\\
A1644&0.12$_{-0.08}^{+0.04}$&14.65$_{-4.17}^{+3.89}$&33.95$_{-3.16}^{+4.12}$&0.232&  2.32&0.75$_{-0.07}^{+0.07}$&3.11\\
A1650&1.44$_{-0.29}^{+0.33}$&1.83$_{-0.37}^{+0.35}$ &81.43$_{-4.81}^{+5.85}$&0.206& 2.06&0.43$_{-0.22}^{+0.12}$&2.22\\
A1651 &0.51$_{-0.17}^{+0.16}$&3.69$_{-0.71}^{+0.74}$&37.81$_{-3.61}^{+3.75}$& 0.194&1.94&0.91$_{-0.08}^{+0.22}$&2.18\\
A1664 &0.29$_{-0.14}^{+0.11}$&6.64$_{-0.81}^{+0.83}$&40.35$_{-3.78}^{+3.71}$&0.176& 1.76&0.72$_{-0.25}^{+0.26}$&2.36\\
A1689 &0.21$_{-0.05}^{+0.05}$&7.15$_{-0.85}^{+0.91}$&39.52$_{-3.92}^{+4.17}$& 0.245&2.45&1.10$_{-0.16}^{+0.18}$&1.54\\
A1736 &0.10$_{-0.01}^{+0.00}$&15.87$_{-4.13}^{+4.16}$&11.26$_{-2.18}^{+1.93}$&0.138& 1.38&1.08$_{-0.27}^{+0.13}$&2.34\\
A1795 &0.43$_{-0.21}^{+0.20}$&4.21$_{-0.88}^{+0.55}$&20.46$_{-2.19}^{+3.16}$&0.105& 1.05&1.04$_{-0.28}^{+0.45}$&2.13\\
A1835 &0.87$_{-0.27}^{+0.18}$&2.36$_{-0.44}^{+0.35}$ &41.93$_{-4.28}^{+5.16}$&0.183& 1.83&0.68$_{-0.15}^{+0.13}$&1.77\\
A1914 &3.05$_{-0.37}^{+0.48}$&0.74$_{-0.33}^{+0.36}$&93.76$_{-10.27}^{+9.61}$&0.121& 1.21&0.43$_{-0.18}^{+0.18}$&1.28\\
A1995 &0.88$_{-0.28}^{+0.33}$&2.77$_{-0.33}^{+0.38}$ &56.22$_{-4.91}^{+6.17}$& 0.124&1.24&0.55$_{-0.18}^{+0.18}$&1.22\\
A2029 &0.37$_{-0.19}^{+0.17}$&6.39$_{-0.43}^{+0.41}$ &42.42$_{-4.19}^{+5.18}$& 0.126&1.26&0.91$_{-0.17}^{+0.37}$&2.20\\
A2034 &0.90$_{-0.22}^{+0.23}$&2.37$_{-0.42}^{+0.49}$ &21.22$_{-2.18}^{+1.93}$&0.106& 1.06&1.06$_{-0.27}^{+0.37}$&2.38\\
A2052 &0.13$_{-0.04}^{+0.03}$&9.63$_{-1.13}^{+1.19}$&13.45$_{-1.28}^{+2.46}$&0.147& 1.47&1.62$_{-0.22}^{+0.27}$&2.32\\
A2063 &0.22$_{-0.05}^{+0.11}$&6.89$_{-0.43}^{+0.48}$&20.06$_{-2.46}^{+2.47}$& 0.194&1.94&0.80$_{-0.28}^{+0.27}$&0.92\\
A2065 &1.22$_{0.18}^{+0.21}$&2.02$_{-0.46}^{+0.34}$&94.05$_{-11.28}^{+10.27}$&0.238& 2.38&0.41$_{-0.27}^{+0.29}$&1.30\\
A2124&0.11$_{-0.02}^{+0.04}$&10.76$_{-1.81}^{+1.84}$ &20.67$_{-2.19}^{+2.15}$&0.157& 1.57&0.91$_{-0.16}^{+0.12}$&2.47\\
A2142&0.50$_{-0.11}^{+0.12}$&5.10$_{-0.49}^{+0.43}$ &21.69$_{-1.28}^{+3.16}$&0.072& 0.72&0.93$_{-0.17}^{+0.12}$&1.29\\
A2147&0.16$_{-0.03}^{+0.02}$&10.06$_{-1.27}^{+1.81}$ &18.09$_{-1.27}^{+2.52}$&0.165& 1.65&1.12$_{-0.19}^{+0.19}$&2.28\\
A2163&2.00$_{-0.45}^{+0.47}$&1.88$_{-0.23}^{+0.27}$&143.51$_{-15.37}^{+14.92}$& 0.129&1.29&0.51$_{-0.15}^{+0.14}$&0.75\\
A2199 &0.24$_{-0.10}^{+0.09}$&  5.98$_{-0.36}^{+0.45}$ &24.25$_{-2.19}^{+2.38}$&0.195& 1.95&0.90$_{-0.10}^{+0.39}$&2.17\\
A2204 &0.76$_{-0.17}^{+0.21}$&  2.56$_{-0.54}^{+0.42}$ &43.67$_{-5.19}^{+5.82}$&0.186& 1.86&0.85$_{-0.37}^{+0.36}$&2.28\\
A2218 &0.21$_{-0.03}^{+0.05}$&  6.13$_{-0.65}^{+1.14}$ &8.95$_{-0.93}^{+2.15}$& 0.138&1.38&1.78$_{-0.15}^{+0.14}$&2.10\\
A2219 &0.66$_{-0.17}^{+0.15}$&  3.24$_{-0.53}^{+0.43}$ &47.06$_{-4.19}^{+5.38}$&0.128& 1.28&0.83$_{-0.33}^{+0.37}$&1.18\\
A2244 &1.12$_{-0.27}^{+0.25}$&  2.23$_{-0.31}^{+0.34}$ &35.20$_{-3.82}^{+4.16}$&0.096& 0.96&0.81$_{-0.17}^{+0.16}$&2.19\\
A2256 &2.67$_{-0.17}^{+0.44}$&  1.06$_{-0.37}^{+0.47}$ &57.77$_{-6.92}^{+5.97}$&0.120& 1.20&1.12$_{-0.27}^{+0.11}$&1.11\\
&&&&&&&\\
\hline
\end{tabular}
\end{table*}

\setcounter{table}{2}
\begin{table*}
\centering
\fontsize{7}{7}
\selectfont
\caption{\fontsize{9}{9} \selectfont Continued.}
\bigskip
\begin{tabular}{lccccccc}
\hline
&&&&&&&\\
\multicolumn{8}{c} {Dark matter profiles}\\
&&&&&&&\\
\multicolumn{6}{c}{NFW profile ($\alpha$=1)} &\multicolumn{2}{c}{NFW ($\alpha$ is free parameter)}\\
Name &r$_{s}$& c$_{200}$& M$_{b}$&$f_{M_{b}/M_{tot}}$ &$\chi^{2}/$d.o.f. &$\alpha$&$\chi^{2}/$d.o.f.\\
     &Mpc& & 10$^{13} M_{\odot}$&& & &\\
&&&&&&&\\
\hline
&&&&&&&\\
A2319 &2.49$_{-0.29}^{+0.22}$&  1.08$_{-0.32}^{+0.37}$ &71.45$_{-6.93}^{+8.14}$&0.155& 1.55&0.86$_{-0.10}^{+0.12}$&2.22\\
A2390 &1.17$_{-0.26}^{+0.34}$&  1.91$_{-0.32}^{+0.42}$ &50.89$_{-4.86}^{+5.67}$&0.180& 1.80&0.94$_{-0.11}^{+0.12}$&2.02\\
A2462 &0.13$_{-0.06}^{+0.03}$&  12.63$_{-0.37}^{+0.42}$ &28.33$_{-3.11}^{+3.14}$&0.136& 1.36&1.21$_{-0.10}^{+0.15}$&2.05\\
A2537 &0.40$_{-0.11}^{+0.12}$&  4.18$_{-0.44}^{+0.34}$ &31.45$_{-3.19}^{+4.28}$&0.195& 1.95&1.18$_{-0.11}^{+0.13}$&2.66\\
A2589 &0.31$_{-0.07}^{+0.09}$&  5.27$_{-0.52}^{+0.45}$ &18.43$_{-2.91}^{+3.15}$&0.183& 1.83&1.34$_{-0.17}^{+0.17}$&2.48\\
A2597 &0.18$_{-0.03}^{+0.05}$&  8.02$_{-0.86}^{+0.77}$ &12.34$_{-1.28}^{+1.27}$&0.114& 1.14&0.92$_{-0.18}^{+0.11}$&1.12\\
A2634 &0.15$_{-0.03}^{+0.05}$&  10.68$_{-1.16}^{+1.17}$ &22.11$_{-2.18}^{+2.63}$&0.186& 1.86&1.21$_{-0.24}^{+0.19}$&2.47\\
A2657 &0.33$_{-0.10}^{+0.11}$&  5.14$_{-0.51}^{+0.43}$ &19.53$_{-2.72}^{+2.15}$&0.197&  1.97&1.18$_{-0.15}^{+0.13}$&2.55\\
A2667 &1.00$_{-0.12}^{+0.23}$&  2.21$_{-0.25}^{+0.47}$ &37.59$_{-3.18}^{+4.72}$&0.190&  1.90&0.74$_{-0.11}^{+0.14}$&2.28\\
A2670 &0.41$_{-0.16}^{+0.13}$&  5.01$_{-0.59}^{+0.51}$ &30.09$_{-3.18}^{+3.61}$&0.180&  1.80&0.74$_{-0.12}^{+0.12}$&1.21\\
A2717 &0.30$_{-0.04}^{+0.05}$&  5.88$_{-0.51}^{+0.54}$ &25.13$_{-2.15}^{+2.34}$&0.194&  1.94&1.03$_{-0.15}^{+0.14}$&2.83\\
A2744 &1.32$_{-0.37}^{_0.30}$&  1.52$_{-0.57}^{+0.35}$ &48.97$_{-4.72}^{+5.16}$&0.220&  2.20&0.70$_{-0.13}^{+0.12}$&2.36\\
A3112 &0.21$_{-0.03}^{+0.04}$&  9.16$_{-0.63}^{+0.43}$ &13.61$_{-1.27}^{+2.17}$&0.072&  0.72&0.83$_{-0.18}^{+0.15}$&1.27\\
A3158 &1.20$_{-0.24}^{+0.13}$&  2.02$_{-0.61}^{+0.60}$ &16.76$_{-1.27}^{+2.15}$& 0.065& 0.65&0.95$_{-0.12}^{+0.14}$&1.22\\
A3266 &0.30$_{-0.05}^{+0.05}$&  8.10$_{-0.51}^{+0.53}$ &42.05$_{-4.17}^{+5.42}$& 0.126& 1.26&1.23$_{-0.17}^{+0.15}$&1.21\\
A3376 &0.32$_{-0.04}^{+0.03}$&  7.01$_{-1.13}^{+1.11}$ &30.67$_{-3.27}^{+4.27}$& 0.115& 1.15&1.63$_{-0.17}^{+0.35}$&2.29\\
A3391 &0.21$_{-0.03}^{+0.05}$&  9.06$_{-0.76}^{+0.54}$ &30.61$_{-3.27}^{+4.28}$& 0.161& 1.61&0.72$_{-0.18}^{+0.19}$&2.02\\
A3395 &0.71$_{-0.20}^{+0.11}$&  3.34$_{-0.47}^{+0.45}$ &45.40$_{-5.28}^{+6.87}$& 0.134& 1.34&0.82$_{-0.13}^{+0.14}$&2.07\\
A3526 &0.10$_{-0.03}^{+0.04}$&  10.13$_{-1.23}^{+1.11}$ &18.24$_{-1.92}^{+2.36}$&0.220& 2.20&0.69$_{-0.13}^{+0.14}$&2.66\\
A3558 &0.22$_{-0.07}^{+0.04}$&  7.89$_{-0.43}^{+0.48}$ &28.64$_{-2.37}^{+4.25}$&0.134&  1.34&0.76$_{-0.28}^{+0.18}$&1.33\\
A3562 &0.22$_{-0.06}^{+0.05}$&  7.96$_{-0.57}^{+0.55}$ &33.07$_{-3.27}^{+4.27}$&0.235&  2.35&0.67$_{-0.19}^{+0.18}$&1.17\\
A3571 &0.25$_{-0.03}^{+0.06}$&  8.52$_{-1.10}^{+0.98}$ &39.35$_{-3.28}^{+4.46}$& 0.129& 1.29&0.91$_{-0.19}^{+0.17}$&2.36\\
A3667 &0.23$_{-0.04}^{+0.06}$&  8.86$_{-0.66}^{+0.65}$ &44.18$_{-4.74}^{+4.56}$& 0.164& 1.64&0.87$_{-0.16}^{+0.17}$&2.22\\
A3827 &0.81$_{-0.23}^{+0.09}$&  4.15$_{-0.57}^{+0.47}$ &70.79$_{-6.48}^{+8.36}$& 0.089& 0.89&0.53$_{-0.34}^{+0.24}$&2.02\\
A3921 &0.46$_{-0.17}^{+0.08}$&   4.87$_{-0.31}^{+0.43}$ &18.56$_{-2.74}^{+2.36}$&0.062& 0.62&1.16$_{-0.29}^{+0.48}$&2.33\\
A4038 &0.15$_{-0.01}^{+0.02}$&  8.78$_{-1.21}^{+1.03}$ &14.98$_{-1.48}^{+2.10}$& 0.180& 1.80&0.77$_{-0.13}^{+0.12}$&2.32\\
A4059 &0.17$_{-0.00}^{+0.02}$&  8.86$_{-1.01}^{+1.02}$ &13.64$_{-1.28}^{+1.25}$& 0.127& 1.27&0.83$_{-0.15}^{+0.12}$&2.37\\
AWM4 & 0.15$_{-0.02}^{+0.03}$& 6.61$_{-1.11}^{+1.32}$ &15.61$_{-1.47}^{+1.47}$&0.238&  2.38&1.52$_{-0.22}^{+0.25}$&2.22\\
CLJ1226.9+3332 &1.54$_{-0.11}^{+0.13}$&  1.54$_{-0.34}^{+0.42}$ &94.12$_{-9.27}^{+10.16}$&0.065& 0.65&0.52$_{-0.11}^{+0.16}$&1.45\\
Coma (A1656) &2.27$_{-0.45}^{+0.33}$&  1.26$_{-0.37}^{+0.35}$ &76.82$_{-4.58}^{+7.62}$&0.183& 1.83&0.63$_{-0.16}^{+0.16}$&1.24\\
IIIZw54 &0.12$_{-0.02}^{+0.01}$ & 10.65$_{-1.28}^{+1.13}$ &17.99$_{-2.15}^{+1.74}$&0.204& 2.04&1.43$_{-0.27}^{+0.28}$&1.16\\
ISCS J1438.1+3414 &6.41$_{-0.42}^{+0.35}$&  0.43$_{-0.08}^{+0.05}$ &316.22$_{-74.83}^{+112.92}$&0.107& 1.07&0.35$_{-0.14}^{+0.14}$&2.73\\
MACSJ0011.7-1523 & 0.65$_{-0.13}^{+0.23}$&3.83$_{-0.23}^{+0.33}$ &59.97$_{-4.28}^{+4.88}$&0.124 &1.24 &0.63$_{-0.18}^{+0.16}$&1.33\\
MACSJ0159.8-0849 & 0.44$_{-0.11}^{+0.11}$&4.85$_{-0.32}^{+0.22}$ &52.57$_{-4.92}^{+5.18}$&0.119 &  1.19 &0.72$_{-0.36}^{+0.24}$&1.19\\
MACSJ0242.6-2132 & 0.27$_{-0.04}^{+0.06}$& 7.18$_{-1.13}^{+1.24}$ &48.48$_{-4.83}^{+4.28}$&0.176 &1.76 &0.64$_{-0.17}^{+0.17}$&2.33\\
MACSJ0429.6-0253 & 0.55$_{-0.11}^{+0.23}$&2.96$_{-1.15}^{+1.36}$ &31.31$_{-3.18}^{+3.24}$&0.169 &1.69 &0.93$_{-0.17}^{+0.17}$&2.38\\
MACSJ0647.7+7015 & 3.12$_{-0.16}^{+0.15}$& 0.90$_{-0.21}^{+0.11}$ &98.25$_{-7.28}^{+8.12}$&0.109& 1.09&0.77$_{-0.27}^{+0.37}$&2.12\\
MACSJ0744.8+3927 &  3.14$_{-0.17}^{+0.19}$&0.81$_{-0.16}^{+0.27}$  &97.81$_{-7.99}^{+9.16}$&0.091& 0.91&0.65$_{-0.19}^{+0.19}$&2.28\\
MACSJ1115.8+0129 & 1.00$_{-0.12}^{+0.19}$& 1.83$_{-0.25}^{+0.33}$  &49.38$_{-5.17}^{+4.76}$&0.209 &2.09 &0.64$_{-0.25}^{+0.26}$&2.21\\
MACSJ1311.0-0311 & 0.50$_{-0.06}^{+0.07}$& 4.51$_{-0.15}^{+0.23}$ &68.15$_{-6.72}^{+7.15}$& 0.123&1.23&0.54$_{-0.13}^{+0.23}$&1.18\\
MACSJ1423.8+2404 & 0.73$_{-0.21}^{+0.19}$& 2.93$_{-0.23}^{+0.35}$ &34.90$_{-3.55}^{+3.26}$& 0.072&0.72&0.63$_{-0.18}^{+0.18}$&2.88\\
MACSJ1427.6-2521 & 0.26$_{-0.02}^{+0.07}$& 7.58$_{-0.77}^{+0.99}$ &34.75$_{-3.33}^{+3.17}$&0.138 &1.38 &0.84$_{-0.23}^{+0.33}$&1.18\\
MACSJ1720.3+3536 & 0.48$_{-0.10}^{+0.08}$& 4.85$_{-0.46}^{+0.31}$ &37.61$_{-4.83}^{+4.29}$&0.092 & 0.92 &0.79$_{-0.17}^{+0.17}$&2.72\\
MACSJ1931.8-2635 & 0.57$_{-0.16}^{+0.15}$& 3.61$_{-0.34}^{+0.26}$ &42.24$_{-6.18}^{+7.11}$&0.129 &1.29 &0.77$_{-0.33}^{+0.38}$&1.44\\ 
MACSJ2129.4-0741 & 1.96$_{-0.37}^{+0.22}$& 1.32$_{-0.21}^{+0.28}$  &95.79$_{-6.19}^{+7.91}$&0.099& 0.99&0.64$_{-0.18}^{+0.18}$&2.46\\
MACSJ2229.8-2756 &0.27$_{-0.03}^{+0.06}$ & 7.64$_{-1.16}^{+0.67}$ &35.64$_{-5.19}^{+4.82}$ &0.077 &0.77 &0.84$_{-0.19}^{+0.19}$&2.33\\
MKW3s & 0.13$_{-0.03}^{+0.03}$ &10.47$_{-1.73}^{+1.18}$ &20.87$_{-4.18}^{+3.99}$&0.173& 1.73&0.85$_{-0.19}^{+0.17}$&2.22\\
MKW4 & 0.05$_{-0.00}^{+0.01}$& 18.93$_{-1.17}^{+1.15}$ &7.69$_{-2.19}^{+1.94}$& 0.346& 3.46&1.75$_{-0.27}^{+0.28}$&2.37\\
MKW8 & 0.10$_{-0.01}^{+0.02}$& 14.68$_{-1.16}^{+1.14}$ &8.34$_{-1.87}^{+2.37}$& 0.111& 1.11&1.83$_{-0.17}^{+0.18}$&0.92\\
PKS0745-191 &0.38$_{-0.02}^{+0.03}$& 5.75$_{-0.41}^{+0.53}$&30.04$_{-2.93}^{+2.84}$&0.089& 0.89&0.88$_{-0.19}^{+0.18}$&0.82\\
RXCJ0043.4-2037 &0.14$_{-0.01}^{+0.03}$&  7.42$_{-1.03}^{+1.89}$ &10.29$_{-3.19}^{+2.55}$&0.161 & 1.61 &1.43$_{-0.18}^{+0.18}$&2.11\\
RXCJ0232.2-4420 &1.27$_{-0.13}^{+0.12}$&  1.58$_{-0.36}^{+0.47}$ &16.51$_{-1.55}^{+1.48}$&0.090 & 0.90 &1.31$_{-0.38}^{+0.34}$&2.22\\
RXCJ0307.0-2840 &0.55$_{-0.16}^{+0.17}$&  2.52$_{-0.37}^{+0.38}$ &21.28$_{-2.11}^{+2.10}$&0.128& 1.68&1.21$_{-0.19}^{+0.19}$&2.38\\
RXCJ0516.7-5430 &0.74$_{-0.16}^{+0.19}$&  2.35$_{-0.67}^{+0.55}$ &17.59$_{-1.99}^{+2.10}$&0.126& 1.26&1.34$_{-0.16}^{+0.17}$&2.72\\
RXCJ0547.6-3152 &0.41$_{-0.07}^{+0.06}$&  3.64$_{-0.53}^{+0.77}$  &18.82$_{-2.18}^{+2.12}$&0.150& 1.70&1.64$_{-0.15}^{+0.15}$&2.34\\
RXCJ0605.8-3518 &0.34$_{-0.02}^{+0.02}$&  3.74$_{-0.33}^{+0.23}$  &13.84$_{-1.28}^{+1.38}$&0.117& 2.17&1.54$_{-0.19}^{+0.19}$&2.38\\
RXCJ1131.9-1955 &0.75$_{-0.13}^{+0.11}$&  2.25$_{-0.67}^{+1.14}$ &20.92$_{-1.28}^{+2.10}$&0.115& 1.35&1.42$_{-0.18}^{+0.18}$&2.88\\
RXCJ2014.8-2430 &0.46$_{-0.04}^{+0.07}$&  3.58$_{-0.47}^{+0.34}$  &18.27$_{-1.93}^{+1.58}$& 0.123&1.73&1.64$_{-0.18}^{+0.19}$&1.83\\
RXCJ2129.6+0005 &0.42$_{-0.06}^{+0.07}$&  3.17$_{-0.34}^{+0.48}$ &12.27$_{-1.26}^{+1.33}$& 0.115&1.45&1.73$_{-0.22}^{+0.24}$&1.33\\

&&&&&&&\\
\hline 
\end{tabular}
\end{table*}

\setcounter{table}{2}
\begin{table*}
\centering
\fontsize{7}{7}
\selectfont
\caption{\fontsize{9}{9} \selectfont Continued.}
\bigskip
\begin{tabular}{lccccccc}
\hline
&&&&&&&\\
\multicolumn{8}{c} {Dark matter profiles}\\
&&&&&&&\\
\multicolumn{6}{c}{NFW profile ($\alpha$=1)} &\multicolumn{2}{c}{NFW ($\alpha$ is free parameter)}\\
Name &r$_{s}$& c$_{200}$&M$_{b}$&$f_{M_{b}/M_{tot}}$ &$\chi^{2}/$d.o.f. &$\alpha$&$\chi^{2}/$d.o.f.\\
     &Mpc& & 10$^{13} M_{\odot}$& && &\\
&&&&&&&\\
\hline
&&&&&&&\\
RXCJ2337.6+0016 &0.29$_{-0.03}^{+0.06}$&  4.72$_{-0.77}^{+0.55}$  &13.75$_{-1.55}^{+1.44}$& 0.131&1.61&1.67$_{-0.19}^{+0.19}$&1.18\\
ZwCL1215 &0.41$_{-0.16}^{+0.08}$&  5.12$_{-0.51}^{+0.42}$ &36.33$_{-2.38}^{+3.17}$& 0.118&1.08&0.74$_{-0.13}^{+0.14}$&1.14\\
RCSJ0224-0002 &0.51$_{-0.17}^{+0.09}$&2.48$_{-0.35}^{+0.35}$&22.45$_{-0.11}^{+0.19}$&0.113 &2.13 &1.02$_{-0.14}^{+0.15}$&1.39\\
RCSJ0439-2904&0.20$_{-0.03}^{+0.01}$ &3.92$_{-0.23}^{+0.35}$&9.01$_{-2.18}^{+1.66}$&0.207&2.37&1.84$_{-0.33}^{+0.36}$&2.20\\
RCSJ1107-0523&0.34$_{-0.03}^{+0.03}$ &2.75$_{-0.35}^{+0.36}$&12.65$_{-1.48}^{+1.47}$&0.219&2.39&1.53$_{-0.53}^{+0.54}$&1.25\\
RCSJ1419.2+5326& 0.22$_{-0.05}^{+0.03}$&  5.74$_{-0.31}^{+0.53}$&33.91$_{-3.17}^{+2.88}$ &0.222 & 2.62 &0.67$_{-0.13}^{+0.12}$&1.28\\
RCSJ1620+2929 &0.31$_{-0.06}^{+0.15}$&3.68$_{-0.41}^{+0.42}$&14.15$_{-1.82}^{+1.47}$&0.147& 1.87&1.04$_{-0.15}^{+0.14}$&1.17\\
RCS2156+0123&1.77$_{-0.34}^{+0.22}$ &1.03$_{-0.33}^{+0.43}$&42.97$_{-5.28}^{+5.82}$&0.181&1.88&0.58$_{-0.12}^{+0.11}$&1.25\\
RCSJ2318+0034&1.65$_{-0.25}^{+0.23}$ &1.08$_{-0.27}^{+0.29}$&51.45$_{-5.82}^{+5.27}$&0.141&1.81&0.64$_{-0.14}^{+0.15}$&1.10\\
RCSJ2319+0038 &0.57$_{-0.09}^{+0.11}$&2.92$_{-0.35}^{+0.36}$&91.91$_{-9.45}^{+8.88}$&0.204&2.24&0.54$_{-0.16}^{+0.15}$&2.36\\
RXJ0439.0+0520& 0.17$_{-0.03}^{+0.05}$& 7.31$_{-0.77}^{+0.27}$&11.43$_{-1.37}^{+2.16}$&0.225 &1.25 &1.04$_{-0.38}^{+0.29}$&2.46\\
RXJ0848.7+4456 &1.33$_{-0.27}^{+0.34}$& 0.57$_{-0.11}^{+0.15}$&9.36$_{-2.19}^{+3.56}$&0.241& 2.81&1.01$_{-0.15}^{+0.14}$&1.66\\
RXJ0849+4452 &1.73$_{-0.25}^{+0.37}$ & 0.74$_{-0.12}^{+0.21}$&67.87$_{-6.84}^{+6.48}$&0.165&1.45&0.55$_{-0.16}^{+0.17}$&1.28\\
RXJ0910+5422 &0.83$_{-0.19}^{+0.15}$& 2.44$_{-0.11}^{+0.15}$&66.84$_{-7.55}^{+6.77}$&0.143&1.03&0.63$_{-0.13}^{+0.13}$&2.46\\
RXJ1113.1-2615 &0.26$_{-0.04}^{+0.06}$&  2.98$_{-0.21}^{+0.23}$&12.72$_{-1.48}^{+1.49}$&0.219& 2.39&1.04$_{-0.37}^{+0.33}$&2.22\\
RXJ1221.4+4918&1.01$_{-0.11}^{+0.15}$&2.02$_{-0.25}^{+0.27}$&75.08$_{-4.29}^{+6.32}$&0.151&1.61&0.54$_{-0.18}^{+0.18}$&1.27\\
&&&&&&&\\
\hline 
\end{tabular}
\end{table*}

Baryons and the baryon diffused component presence produces in general a flattening of the density profile (\citealt{El-Zant:01}, \citealt{El-Zant:04}; \citealt{Romano-Diaz:08}, \citealt{delPopolo:09}, \citealt{Governato:10}), and rounder halos than those seen in DM simulations (\citealt{Debattista:08}; \citealt{Abadi:10}). The final configuration of a cluster is fixed by the initial quantity of baryons present in the proto cluster and by collapse/formation process. Then it is logic to expect that the final central baryonic content and the BCG mass is somehow correlated with baryonic and total cluster mass, $M_{cl}$. For example, \citet{Whiley:08}, found that $M_{BCG}\propto M_{cl}^{0.4}$ or $M_{cl}^{0.5}$ depending on the feedback model used. Also a correlation between BCG luminosity and cluster X-ray luminosity was found by several authors (\citealt{Schombert:88}; \citealt{Edge:91}; \citealt{Edge:91}; \citealt{Hudson:97}). \citet{Whiley:08} measured the quoted correlation as $M_{BCG}\propto M_{cl}^{0.12\pm0.03}$ for K~band magnitudes inside a diameter of 37~kpc (radius of 13$h^{-1}$~kpc). \citet{Brough:08} found $L_{BCG}\propto M_{cl}^{0.11\pm0.10}$ at K~band inside 12$h^{-1}$~kpc (several other results are given in \citealt{Lin:04}; \citealt{Popesso:07}; \citealt{Yang:08}; \citealt{Haarsma:10}; Fedeli 2012). 
The flattening of the slope can be interpreted as due to the fact that the presence of a larger quantity of baryons guarantees
a larger transfer of energy and angular momentum from baryons to DM, with the result that DM moves to larger orbits, reducing the
inner density (\citealt{delPopolo:09}, \citealt{delPopolo:12}; \citealt{El-Zant:01}, \citealt{El-Zant:04}; \citealt{Romano-Diaz:08}, \citealt{Romano-Diaz:09}).  

As already reported, NFW N-body simulations predicted a profile with inner slope $\propto r^{-1}$ and an outer one $\propto r^{-3}$. More recently, it has been shown that the density profiles are better fitted by an Einasto profile which becomes shallower towards the centre of the halo (e.g. \citealt{Navarro:10}). In any case, the lower values for the inner slope obtained in N-body simulations are $\alpha=0.8$ at 120 pc (\citealt{Stadel:09}) in agreement to analytical and numerical works to solve the Jeans equation (\citealt{Austin:05}; \citealt{Dehnen:05}). Inner density slopes in the range $0.7 <\alpha< 1.5$ could be considered in agreement with $\Lambda$CDM predictions (SA07). 

However, as reported in the introduction, slopes flatter than those obtained in the simulations, $\alpha< 0.7$ are observed in some clusters (\citealt{Sand:02}, \citealt{Sand:04}, \citealt{Sand:08}; \citealt{Newman:09}, \citealt{Newman:11}) and the largest numbers of dwarf galaxies and LSBs, and, as previously reported, also some of our clusters have slopes $\alpha <0.7$.  

This discrepancy does not necessarily imply a problem for the $\Lambda$CDM model (e.g. see DP09; \citealt{Governato:10}). 

Given the several and noteworthy pieces of evidence supporting $\Lambda$CDM on large scales, the discrepancy could be connected to 
the fact that baryonic physics is a fundamental issue in cluster formation. It could originate just because we are comparing dissipationless systems generated by N-body simulations with real, dissipational structures, whose physics is different from the dissipationless physics typical of DM. We should not forget that the inner 10 kpc of clusters are dominated by baryons (e.g., \citealt{Sand:02}, \citealt{Sand:04}, \citealt{Sand:08}; \citealt{Newman:09}, \citealt{Newman:11}) whose presence strongly influence the DM distribution. While baryons can steepen the inner slope of the density profile of clusters through the adiabatic contraction of DM (\citealt{Blumenthal:86}; \citealt{Gnedin:04}; \citealt{Gustafsson:06}), heating of DM due to dynamical friction with cluster galaxies  can counteract the adiabatic contraction effect and flatten the inner profile (\citealt{El-Zant:01}, \citealt{El-Zant:04}; \citealt{Romano-Diaz:08}).

\section{Discussion}

As discussed in the paper, our analysis used Chandra data and similar methods to that of SA07, but our sample is larger.
Confronting our tables and those of SA07, we see several clusters common to the two studies (e.g., A1795; A2029; A478; A1413; A2204; A383; A963; A1835; A611; A2537; MACSJ0242.6-2132; MACSJ1427.6-2521; MACSJ2229.8-2756; MACSJ1391.8-2635; MACSJ1115.8+0129; MACSJ1720.3+3536; MACSJ1311.0-0311; MACSJ1423.8+2404). The inner slopes of these clusters are compatible in our and SA07 study. 

Other clusters, like A383, A611, A963, A1835, A2029, A2204, A2589 were studied by others authors. For example, A611 was studied by \citet{Newman:09} who combined weak lensing from multicolor Subaru imaging, strong lensing (Hubble Space Telescope) and stellar velocity dispersion measures (Keck Telescope), sampling the dark matter profile from 3 kpc to 3.25 Mpc. Newman et al. (2009) found values of $r_s = 320^{+240}_{-110}$ kpc and $c = 5.1^{+1.7}_{-1.6}$ (in agreement with SA07), but $\alpha < 0.3$($<0.56$, $<0.65$) at 68 per cent (95 per cent, 99 per cent) CL. SA07 found a value of the slope $\alpha = 0.64^{+0.94}$. A383 was studied by S08 finding a flat inner DM slope ($\alpha = 0.45^{+0.2}_{-0.25}$), and by Newman et al. (2011) who found $\alpha < 1$ at (95 per cent confidence) and a best fit (inferred from weak and strong lensing, kinematics and X-ray data) of $\alpha = 0.59^{+0.30}_{-0.35}$. SA07 found $\alpha < 0.8$, we found $\alpha = 1.58^{+0.22}_{-0.22}$.

A flat slope was obtained also for A963 by \citet{Sand:08}, but \citet{Bartelmann:04}, \citet{Dalal:03} and SA07 found values consistent with the NFW model. We found $\alpha=1.52^{0.13}_{-0.12}$. In the case of A2029, \citet{Lewis:03} found $\alpha=1.19 \pm 0.04$ similarly to SA07, and we got $0.91^{+0.37}_{-0.17}$. In the case of A1835, A2029, A2204, \citet{Arabadjis:04} by means of Chandra data obtained $\alpha \simeq 0.9$, $\alpha \simeq 1.85$, and $\alpha \simeq 1.8$, respectively, much larger than SA07 values and ours.   

As reported, \citet{Newman:09}, \citet{Newman:11}, presented a detailed analysis of DM and baryonic distributions in A611, and A383 combining weak lensing, strong lensing and stellar velocity dispersion for the BCG, and finding slopes flatter than the NFW predictions. According to \citet{Newman:09}, \citet{Newman:11}, degeneracies in constraining the DM profile can be broken only simultaneously using the three techniques. In reality, the X-ray observations alone give information on clusters on cluster structure in the range 500-50 kpc. At smaller radii, temperature determination is limited by instrumental resolution, substructure (SA07), and they are also limited by ``cooling flows" presence and the breaking of assumption of hydrostatic equilibrium (see \citealt{Arabadjis:04}). 

So, if the \citet{Newman:09} and \citet{Newman:11} point of view is correct, for clusters containing larger quantity of baryons the inner profile may be flatter than X-ray observations.  SA07 tried to understand how important was the role of central BCG in MS2137.3-2353 (for which \citet{Sand:02}, \citet{Sand:04} found a flat profile). Varying the ratio $M/L_{V}$ they observed, in agreement with \citet{Newman:09} and \citet{Newman:11} a flattening of the profile, but in any case the slope was still compatible with a NFW fit (differently from \citet{Newman:09}, \citet{Newman:11}). 

Other systematic uncertainties that could change the results is the presence of non-thermal pressure support (due to gas motions, cosmic rays or magnetic fields), however for relaxed clusters, bulk and/or turbulent motions, if present, could cause changes in the mass measurements of 10-20 per cent accuracy (\citealt{Nagai:07}; \citealt{Rasia:06}). Similar effects are expected by magnetic pressure \citet{Dolag:00}. 

Another important issue to recall is the fact that in the present paper, as in that of SA07 the X-ray data only extend to about $r_{2500}$ (approximately a quarter to a third of the virial radius in most clusters), so, for example, the claim that the mass-concentration relation is appropriate for the virial radii, is not totally true. 

\section{Conclusions}

We presented the reconstruction of the total mass (dark matter, gas and luminous matter) from the Chandra observations of 129 massive X-ray luminous galaxy clusters in the redshift range 0.01 -- 1.4. We estimated the total ($M_{tot}$) mass within $R_{200}$ in the range 3 -- 293 $\times 10^{14} M_{\odot}$.

In order to estimate the fraction of dark matter and gas we have used the conditions of hydrostatic equilibrium and spherical symmetry. Similarly to SA07, we performed two different analysis: the first concerning the total mass of the clusters, and the second decomposing the total mass in its diffuse gas, DM, and BCG components. 

Similarly to SA07, the NFW gives a good fit to the total mass and DM distribution. We obtained a best-fitting result for the inner slope of the DM profile in the clusters $\alpha = 0.94 \pm 0.13$. We also obtained a mass-concentration relation  $c \propto M^a/(1+z)^b$, with $a = $ -0.56 $\pm$ 0.15 and $b = $0.80$ \pm$ 0.25 (68 per cent confidence limits) in agreement with previous results and simulations. 

Finally, we showed that there is a tight correlation among the inner slope $\alpha$ and the baryonic mass, $M_b$, content.

\section*{Acknowledgments}

This research has made use of data obtained from the Chandra Data Archive and the Chandra Source Catalog, and software provided by the Chandra X-ray Center (CXC) in the application packages CIAO, ChIPS, and Sherpa. We thank all the staff members involved in the Chandra project. Iu. Babyk and I. Vavilova note that this work was partially supported in frame of the "CosmoMicroPhysics" Program and Target Project of the Physics and Astronomy Division of the NAS of Ukraine.

\let\jnlstyle=\rm\def\jref#1{{\jnlstyle#1}}\def\aj{\jref{AJ}}
  \def\araa{\jref{ARA\&A}} \def\apj{\jref{ApJ}\ } \def\apjl{\jref{ApJ}\ }
  \def\apjs{\jref{ApJS}} \def\ao{\jref{Appl.~Opt.}} \def\apss{\jref{Ap\&SS}}
  \def\aap{\jref{A\&A}} \def\aapr{\jref{A\&A~Rev.}} \def\aaps{\jref{A\&AS}}
  \def\azh{\jref{AZh}} \def\baas{\jref{BAAS}} \def\jrasc{\jref{JRASC}}
  \def\memras{\jref{MmRAS}} \def\mnras{\jref{MNRAS}\ }
  \def\pra{\jref{Phys.~Rev.~A}\ } \def\prb{\jref{Phys.~Rev.~B}\ }
  \def\prc{\jref{Phys.~Rev.~C}\ } \def\prd{\jref{Phys.~Rev.~D}\ }
  \def\pre{\jref{Phys.~Rev.~E}} \def\prl{\jref{Phys.~Rev.~Lett.}}
  \def\pasp{\jref{PASP}} \def\pasj{\jref{PASJ}} \def\qjras{\jref{QJRAS}}
  \def\skytel{\jref{S\&T}} \def\solphys{\jref{Sol.~Phys.}}
  \def\sovast{\jref{Soviet~Ast.}} \def\ssr{\jref{Space~Sci.~Rev.}}
  \def\zap{\jref{ZAp}} \def\nat{\jref{Nature}\ } \def\iaucirc{\jref{IAU~Circ.}}
  \def\aplett{\jref{Astrophys.~Lett.}}
  \def\apspr{\jref{Astrophys.~Space~Phys.~Res.}}
  \def\bain{\jref{Bull.~Astron.~Inst.~Netherlands}}
  \def\fcp{\jref{Fund.~Cosmic~Phys.}} \def\gca{\jref{Geochim.~Cosmochim.~Acta}}
  \def\grl{\jref{Geophys.~Res.~Lett.}} \def\jcp{\jref{J.~Chem.~Phys.}}
  \def\jgr{\jref{J.~Geophys.~Res.}}
  \def\jqsrt{\jref{J.~Quant.~Spec.~Radiat.~Transf.}}
  \def\memsai{\jref{Mem.~Soc.~Astron.~Italiana}}
  \def\nphysa{\jref{Nucl.~Phys.~A}} \def\physrep{\jref{Phys.~Rep.}}
  \def\physscr{\jref{Phys.~Scr}} \def\planss{\jref{Planet.~Space~Sci.}}
  \def\procspie{\jref{Proc.~SPIE}} \let\astap=\aap \let\apjlett=\apjl
  \let\apjsupp=\apjs \let\applopt=\ao

\bsp
\label{lastpage}
\end{document}